\documentclass[]{aa} 
\usepackage{graphicx}
\usepackage[figuresright]{rotating}
\usepackage{aalongtable}
\usepackage{caption2}
\usepackage{txfonts}
\begin{document}

   \title{Damped and sub-damped Lyman-$\alpha$ absorbers in z $>$ 4 QSOs \thanks{The observations reported here were obtained with the
W. M. Keck Observatory, which is operated by the California Association for
Research in Astronomy, a scientific partnership among the California
Institute of Technology, the University of California, and the National
Aeronautics and Space Administration.}}

   \author{R. Guimar\~aes\inst{1}
          \and
          P. Petitjean\inst{2}
	  \and
	  R. R. de Carvalho\inst{3}
	  \and 
          S.G. Djorgovski\inst{4}
	  \and
	  P. Noterdaeme\inst{5}
	  \and
	  S. Castro\inst{6}
	  \and
	  P. C. da R. Poppe\inst{1}
      \and
          A. Aghaee\inst{7,8}
	  }

   \offprints{R. Guimar\~aes, \email{ rguimara@eso.org}}

   \institute{Universidade Estadual de Feira de Santana - Av. Transnordestina, s/n, 40036-900, Feira de Santana, BA - Brasil
	 \and 
            UPMC Paris 6, Institut d'Astrophysique de Paris, CNRS, 98bis Boulevard Arago - 75014 Paris, FRANCE
     \and
	        Instituto Nacional de Pesquisas Espaciais - INPE, Av. dos Astronautas, 1758, 12227-010, S. J. dos Campos, SP - Brasil
	 \and
             California Institute of Technology, MS 105-24, Pasadena, CA 91125
     \and
            Inter-University Centre for Astonomy and Astrophysics, Post Bag 4, Ganeshkhind, Pune 411 007, India
     \and
	     Europeen Southern Observatory, Karl-Schwarzschild Strasse, 2, Garching, Germany
     \and
             Department of Physics, University of Sistan and Baluchestan, 98135 Zahedan, Iran
	 \and
	     School of Astronomy and Astrophysics, Institute for Research in Fundamental Sciences (IPM), P.O.BOX 193595-5531, Tehran, Iran
             }

   \date{Received 17/12/2008; accepted 07/09/2009}

\abstract{We present the results of a survey for damped (DLA, log~$N$(H~{\sc i})~$>$~20.3) and 
sub-damped Lyman-$\alpha$ systems (19.5~$<$~log~$N$(H~{\sc i})~$<$~20.3) at $z>2.55$ along 
the lines-of-sight to 77 quasars with emission redshifts in the range $4<z_{\rm em}<6.3$. 
Intermediate resolution ($R\sim4300$) spectra have been obtained with the Echellette 
Spectrograph and Imager (ESI)  mounted on the Keck telescope. A total of 100 systems 
with log~$N$(H~{\sc i})~$>$~19.5 are detected of which 40 systems
are damped Lyman-$\alpha$ systems for an absorption length of $\Delta X$~=~378. 
About half of the lines of sight of this homogeneous survey 
have never been investigated for DLAs.
We study the evolution with redshift of the cosmological density of the neutral gas and
find, consistently with previous studies at similar resolution, that 
$\Omega_{\rm DLA, H_I}$ decreases at $z>3.5$. The overall cosmological
evolution of $\Omega_{\rm H_I}$ shows a peak around this redshift. 
The H~{\sc i} column density distribution for log~$N$(H~{\sc i})~$\geq$~20.3 
is fitted, consistently with previous surveys, with a single power-law of index 
$\alpha$~$\sim$~-1.8$\pm$0.25. This power-law overpredicts data at the high-end
and a second, much steeper, power-law (or a gamma function) is needed. There is a flattening 
of the function at lower H~{\sc i} column densities with an index of $\alpha$~$\sim$~$-$1.4 
for the column density range log~$N$(H~{\sc i})~=~19.5$-$21. The fraction of H~{\sc i} mass 
in sub-DLAs is of the order of 30\%. The H~{\sc i} column density distribution does not evolve 
strongly from $z\sim 2.5$ to $z\sim 4.5$. 
	}
\vspace{1pc}

\maketitle

\section{Introduction}
The amount of neutral gas in the Universe is an important ingredient of galaxy formation scenarios
because the neutral phase of the intergalactic medium is the reservoir for star-formation 
activity in the densest places of the universe where galaxies are to be formed. It is therefore 
very important to make a census of the mass in this phase and to determine its cosmological evolution
(see e.g. P\'eroux et al. 2001).

The gas with highest H~{\sc i} column density is detected through damped Lyman-$\alpha$ absorptions
in the spectra of remote quasars. Although damped wings are seen for column densities of the order 
of log~$N$(H~{\sc i})~$\sim$~18, the neutral phase corresponds to log~$N$(H~{\sc i})~$\geq$~19.5 
(Viegas 1995). The column density defining the so-called damped Lyman-$\alpha$ (DLA) systems 
has been taken to be log~$N$(H~{\sc i})~$\geq$~20.3 because this corresponds to the critical 
mass surface density limit for star formation (Wolfe et al. 1986) but also because the equivalent 
width of the corresponding absorption is appropriate for a search of these systems in low 
resolution spectra. Therefore several definitions have been introduced. $\Omega_{\rm g}^{\rm DLA}$ 
is the mass density of baryons in DLA systems, defined arbitrarily as systems with 
log~$N$(H~{\sc i})~$\geq$~20.3. $\Omega_{\rm g}^{\rm H_I}$ is the mass density of neutral hydrogen 
in all systems : DLAs, Lyman limit systems (LLS) and the Lyman-$\alpha$ forest. The mass density of 
H~{\sc i} in the Lyman-$\alpha$ forest is negligible because the slope of the H~{\sc i} column 
density distribution is larger than $-2$ ($\sim-1.5$; the gas is highly ionized). It is more 
difficult to estimate the contribution of LLS as the column density of these systems is 
very difficult to derive directly because the Lyman-$\alpha$ line lies in the logarithmic 
regime of the curve of growth. 

However, $\Omega_{\rm g}^{\rm H_I}$ is not easily related to physical quantities as
the LLS with log~$N$~$<$~19.5 are at least partly ionized when the ones with log~$N$~$>$~19.5 are not
(see e.g. Meiring et al. 2008). 
On the contrary, as emphasized by Prochaska et al. (2005), hereafter PHW05, the mass density 
of the neutral phase,  $\Omega_{\rm g}^{\rm neut}$, is a good indicator of the mass available 
for star-formation and should be prefered instead. 
Note that $\Omega_{\rm g}^{\rm neut}$ is not equal to $\Omega_{\rm g}^{\rm DLA}$. 
The column density limit at which the gas is mostly neutral cannot be defined
precisely but should lie between log~$N$(H~{\sc i})~=~19 and 19.5. In any case, a 
conservative position is to consider that all systems above 19.5 are neutral.

Whether or not the mass of the neutral gas in the systems with 19.5~$<$~log~$N$(H~{\sc i})~$<$~20.3
(the so-called sub-DLAs or super-LLS) is negligible has been the source of intense discussions in recent years.
Note that these discussions are related to the mass in the {\sl neutral} phase only. Indeed, it is
known for long (e.g. Petitjean et al. 1993) that the {\sl total} mass associated with the
Lyman limit systems is larger than that of DLAs. Indeed the gas in the LLS phase is mostly ionized
and located in extended halos whereas DLAs are located in dense and compact regions. 
P\'eroux et al. (2003), hereafter PMSI03, have been the first to consider the sub-DLAs as an important
reservoir of neutral gas. They claim that at $z>3.5$, DLAs could contain only 50\% of the neutral
gas, the rest being to be found in sub-DLAs. When correcting for this, they find that
the comoving mass density shows no evidence for a decrease above $z=2$. 
PHW05 questionned this estimate. They use the Sloan Digital Sky Survey to 
measure the mass density of predominantly neutral gas $\Omega_{\rm g}^{\rm neut}$. They 
find that DLAs contribute $>$80\% of  $\Omega_{\rm g}^{\rm neut}$ at all redshift. 
Uncertainties are very large however and the same authors estimate that the systems with 
log~$N$(H~{\sc i})~$>$~19 (the super-LLS) could contribute 20-50~\% of $\Omega_{\rm g}^{\rm H_I}$. 
Therefore, the question of what is the contribution of super-LLS to $\Omega_{\rm g}^{\rm neut}$
is not settled yet.

In addition, the evolution of $\Omega_{\rm g}^{\rm neut}$ at the very high redshift, $z>4$,
is not known yet. PHW05 claim that there is no evolution of $\Omega_{\rm g}^{\rm DLA}$ for 
$z>3.5$ but they caution the reader that results for $z>4$ should be confirmed with
higher resolution data. The reason is that the Lyman-$\alpha$ forest is so dense
at these redshifts that it is very easy to misidentify a strong blend of lines
with a DLA. Therefore $\Omega_{\rm g}^{\rm DLA}$ can be easily overestimated.

In this paper we present the result of a survey for DLAs and sub-DLAs at high redshift ($z>2.55$)
using intermediate resolution data. We identify a total of 100 systems with log~$N$(H~{\sc i})~$\leq$~19.5
of which 40 are DLAs over the redshift range $2.88 \leq z_{abs} \leq 4.74$ along 77 lines-of-sight towards
quasars with emission redshift 4~$\leq$~$z_{\rm em}$~$\leq$~6.3. 
The sample and data reduction are presented in Section~2. In Section~3 we describe the procedures used to 
select the absorption systems. Section~4 analyses statistical quantitities characterizing the evolution 
of DLAs and sub-DLAs and discusses the cosmological evolution of the neutral gas mass density.  
Conclusions are summarized in Section 5. Throughout the paper, we adopt 
$\Omega_{\rm m} = 0.3$, $\Omega_{\Lambda}=0.7$ and $H_0$ = 72~km~s$^{-1}$.

\section{Observation and data reduction }

Medium resolution (R~$\sim$~4300) spectra of all $z>3$ quasars discovered in the course of the DPOSS
survey (Digital Palomar Observatory Sky Survey; see, e.g., Kennefick et al. 1995, Djorgovski et al. 1999 
and the complete listing of QSOs available at
http://www.astro.caltech.edu/$\sim$george/z4.qsos) have been 
obtained with the Echellette Spectrograph and Imager (ESI, Sheinis et al 2002) mounted on 
the KECK II 10~m telescope. In total, 99 quasars have been observed, 57 of which already reported in the literature (see Table \ref{logbook} for details).
We provide in Table \ref{logbook} a summary of the observation log for the 99 quasars. 
Columns 1 to 8 give, respectively, the quasar's name, the emission redshift, the apparent R magnitude,
the J2000 quasar coordinates, the date of observation, the exposure time and the notes.  

The echelle mode allows to cover the full wavelength range from 3900~\AA~ to 10900~\AA~ in ten orders 
with $\sim 300$~\AA~ overlap between two adjacent orders.
The instrument has a spectral dispersion of about 11.4~km~s$^{-1}$~pixel$^{-1}$ and a pixel size ranging from 
0.16~\AA~pixel$^{-1}$ in the blue to 0.38~\AA~pixel$^{-1}$ in the red. 
The 1~arcsec wide slit is projected onto 6.5~pixel resulting in a $R\sim$4300 spectral resolution.

Data reduction followed standard procedures using IRAF for 70\% of the sample and the programme
{\it makee} for the remaining.  
For the IRAF reduction, the procedure was as follows. The
images were overscan corrected for the dual-amplifier mode. Each amplifier has a different 
baseline value and different gain which were corrected by using a script adapted from LRIS 
called {\it esibias}. Then all the images were bias subtracted and corrected for bad pixels. 
The images were divided by a normalized two-dimensional flat-field image to remove individual pixel 
sensitivity variations. The flat-field image was normalized by fitting its intensity along the dispersion 
direction using a high order polynomial fit, while setting all points outside the order aperture to unity. 
The echelle orders were traced using the spectrum of a bright star.
Cosmic rays were removed from all two-dimensional images.  
For each exposure the quasar spectrum was optimally extracted and background subtracted. 
The task {\it apall} in the IRAF package {\it echelle} was used to do this. 
The CuAr lamps were individually extracted using the quasar's apertures. Lines were identified 
in the arc lamp spectra by using the task {\it ecidentify} and a polynomial was fitted to the 
line positions resulting in a dispersion solution with a mean RMS of 0.09~\AA. 
The dispersion solution computed on the lamp was then assigned to the object spectra 
by using the task {\it dispcor}. Wavelengths and redshifts were computed in the heliocentric 
restframe. The different orders of the spectra were combined using the task {\it scombine} in the IRAF 
package. It is important to note that the signal-to-noise ratio drops sharply at
the edges of the orders. We have therefore carefully controlled this procedure to
avoid any spurious feature. 

The SNR per pixel was obtained in the regions of the Lyman-$\alpha$ forest that are free of absorption 
and the mean SNR value, averaged between the Lyman-$\alpha$ and Lyman-$\beta$ QSO emission lines, was computed. 
We used only spectra with mean SNR $\geq 10$. For simplicity, we excluded from our analysis broad absorption 
line (BAL) quasars. We therefore used 77 lines-of-sight out of the 99 available to us.

The continuum was automatically fitted (Aracil et al. 2004; Guimar\~aes et al. 2007) and the 
spectra were normalized. We checked the normalization for all lines-of-sight and manually corrected for
local defects especially in the vicinity of the Lyman-$\alpha$ emission lines and 
when the Lyman-$\alpha$ forest is strongly blended.

Metallicities measured for twelve $z>3$ DLAs observed along five lines-of-sight of this sample
have been published by Prochaska et al. (2003b); see also Prochaska et al. (2003a).

\section{Identification of Damped and sub-Damped Lyman-$\alpha$ systems}

We used an automatic version of the Voigt profile fitting routine VPFIT (Carswell et al. 1987)
to decompose the Lyman-$\alpha$ forest of the spectra in individual components. As usual, we
restricted our search outside of 3000~km~s$^{-1}$ from the QSO emission redshift. This is to
avoid contamination of the study by proximate effects such as the presence of overdensities
around quasars (e.g. Rollinde et al. 2005, Guimar\~aes et al. 2007). 

From this fit we could identify the candidates with log~$N$(H~{\sc i})~$-$~error~$\geq$~19.5. 
We then carefully inspected each of these candidates individually with special attention to
the following characteristics of DLA absorption lines:

\begin{itemize}
\item the wavelength range over which the line is going to zero;
\item the presence of damped wings;
\item the identification of associated metal lines when possible.
\end{itemize}
The redshift of the H~{\sc i} absorptions was adjusted carefully using the associated metal absorption features
and the final H~{\sc i} column density was then refitted using the high order lines in the Lyman series
when Lyman series are covered by our spectrum and not blended with other lines.
After applying the above criteria, 100 DLAs and sub-DLAs with log~$N$(H~{\sc i})~$\geq$~19.5 were confirmed. 
A total of 65 systems were not previously used for $\Omega_{\rm DLA, H_I} / \Omega_{\rm sub-DLA, H_I}$ estimations, 
21 of which are DLAs and 44 are sub-DLAs. Their
characteristics are given in Table \ref{candidatosdlas}: Columns 1 to 7 give, respectively, 
the QSO's name; the emission redshift estimated as the average of the determinations from the peak 
of the Ly$\alpha$ emission and from the peak of a Gaussian fitted to the CIV emission line; the minimum redshift along a DLA/sub-DLA could be detected; the maximum redshift along a DLA/sub-DLA could be detected; the redshift of the DLA/sub-DLA; the DLA/sub-DLA H~{\sc i} column density; associated detected metal lines and notes. Lyman series and selected associated metal lines are shown in the Appendix "List of Figures". The QSO lines of sight of our sample along which we detect no DLA and/or 
sub-DLA are listed in Table \ref{wbas}. Comments on DLA/sub-DLA systems differences between our measurements and measurements by others are in the Appendix "Notes on Individual Systems". 

The metal lines have been searched for using a search list of the strongest atomic transitions given 
in Table~\ref{princ_lines}.
The corresponding absorptions have been fitted using the package VPFIT. 
A full account of this metallic column densities and the corresponding abundances is out of the
scope of this paper and will be presented elsewhere (Guimar\~aes et al., in preparation).

\section{Analysis}
\begin{figure}[!htb] 
\begin{center}
\includegraphics[height=9cm,width=9cm]{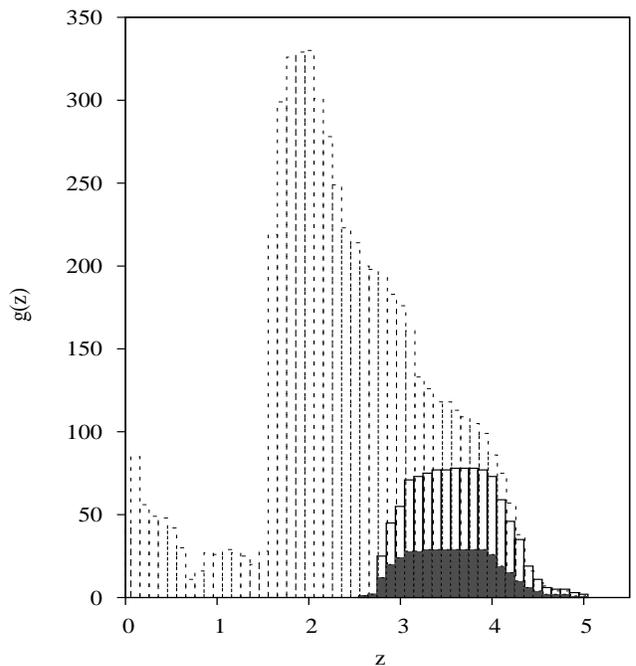}
\caption{Solid curve shows the redshift sensitivity function of our survey for all lines-of-sight, the filled grey histogram is only for unpublished lines-of-sight. Dashed curve shows the redshift sensitivity function of previous surveys computed by PMSI03.}
\label{gz}
\end{center}
\end{figure}

\begin{figure}[!htb] 
\begin{center}
\includegraphics[height=9cm,width=9cm]{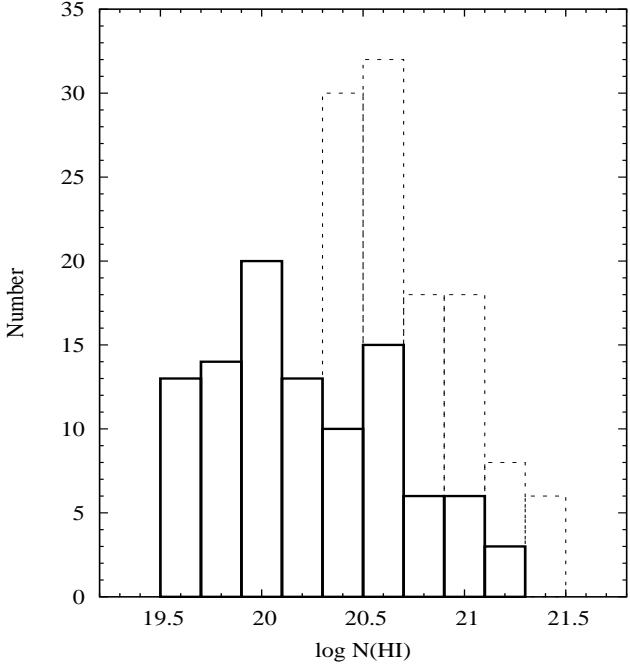}
\caption{Histogram of the H~{\sc i} column densities measured for the 100 DLAs and sub-DLAs with
log~$N$(H~{\sc i})~$\geq$~19.5 detected in our survey (solid-line histogram). The dashed-line histogram represents the H~{\sc i} column densities measured  by PMSI03 with log~$N$(H~{\sc i})~$\geq$~20.3.} 
\label{histHI}
\end{center}
\end{figure}

Using the procedure described in the previous Section, we detect 100 systems with log~$N$~(H~{\sc i})~$\geq$~19.5,
out of which 40 are DLAs. We use this sample to investigate the characteristics of the neutral phase 
over the redshift range 2.5~$\leq$~$z$~$\leq$~5.
We compare in Fig~\ref{gz} the redshift sensitivity function of our survey with the redshift 
sensitivity function of previous surveys computed by PMSI03. 
Although the redshift path of our survey is much smaller than that of the SDSS survey (PHW05), it is
at $z>3.5$ similar to the surveys by P\'eroux et al. (2001, 2003). Note that our survey is
homogeneous and at spectral resolution twice or more larger than previous surveys.
The H~{\sc i} column density distribution of the 100 DLAs and sub-DLAs measured in 
this work and that of the systems with log~$N$(H~{\sc i})~$\geq$~20.3 in PMSI03
are shown in Fig.~\ref{histHI}.
We are confident that we do not miss a large number of sub-DLAs down to the above limit.

\begin{figure}[!htb] 
\begin{center}
\includegraphics[height=9cm,width=9cm]{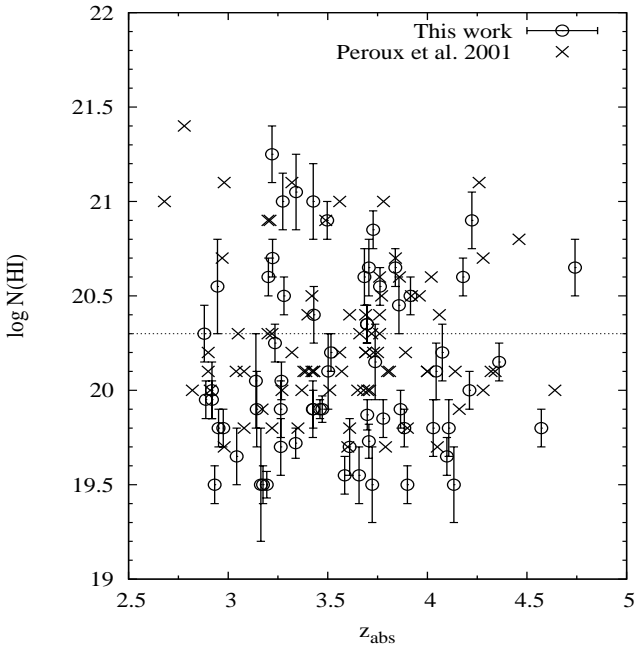}
\caption{The logarithm of the H~{\sc i} column density measured in our unpublished systems (circles)
is plotted versus the redshift. The data points of P\'eroux et al. (2001) are shown
for comparison with crosses.
}
\label{NHIzabs}
\end{center}
\end{figure}

To give a global overview of the survey, we plot in Fig.~\ref{NHIzabs}, log~$N$(H~{\sc i}) 
versus $z_{\rm abs}$ for the 65 unpublished damped/sub-DLA absorption systems. 
In the same figure we show for comparison the data points from the P\'eroux et al. (2001) survey.

\begin{figure}[!htb] 
\begin{center}
\includegraphics[height=9cm,width=9cm]{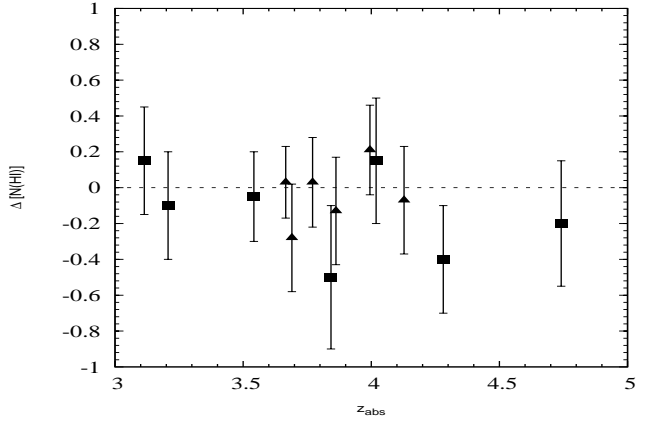}
\caption{The H~{\sc i} column density difference between our measurements and 
those by either P\'eroux et al. (2005) 
(triangles) or Prochaska \& Wolfe (2009) (squares) in systems common to different surveys
is plotted versus redshift. 
}
\label{delta}
\end{center}
\end{figure}

In Figure \ref{delta}, we plot for comparison, as a function of redshift, the 
difference between the H~{\sc i} column densities measured for the same systems 
by us and by either P\'eroux et al. (2005) from high-resolution data or
Prochaska \& Wolfe (2009) from SDSS data. The measurements are consistent within errors.
%

\subsection{Column Density Distribution Function}

\begin{figure} 
\begin{center}
\includegraphics[height=9cm,width=9cm]{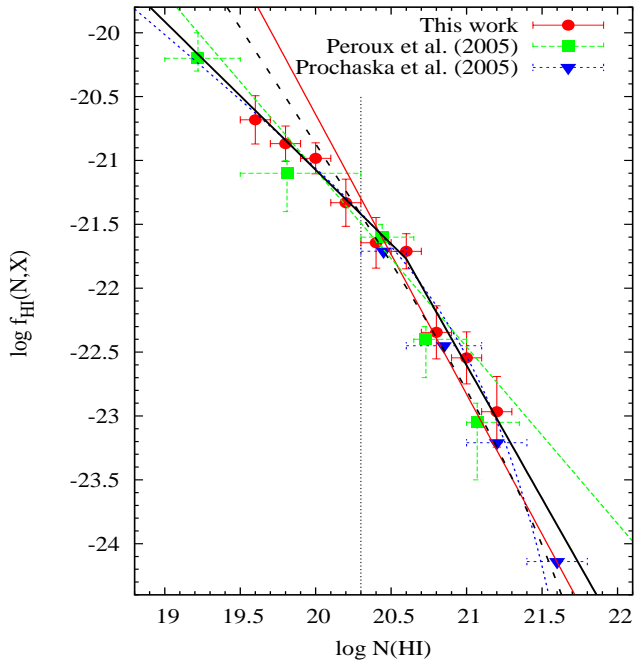}
\caption{Frequency distribution function over the redshift range  $z = 2.55 - 5.03$. 
The dashed green, solide black and dotted blue lines are, respectively, a power-law, a double power-law
and a gamma function fits to the data. The solid red and dashed black lines are, respectively,
a power-law and gamma function fits to the data obtained by PHW05.}
\label{frequency}
\end{center}
\end{figure}

The H~{\sc i} absorption system frequency distribution function is defined as:

\begin{equation}
f(N,X) dN dX = \frac {m_{\rm sys}} {\Delta N \times \sum_i^n \Delta X_{\rm i}} dN dX
\end{equation}

where $m_{\rm sys}$ is the number of absorption systems with a column density comprized
between $N-\Delta N/2$ and $N+\Delta N/2$ and observed over an absorption 
distance interval of $\Delta X$. 
The total absorption distance coverage, 
$\sum_i^n \Delta X_{\rm i}$, is computed over the whole sample of $n$ QSO lines-of-sight. 
The absorption distance, $X$, is defined as

\begin{eqnarray}
X(z) = \int_0^z (1+z)^2 E(z)  dz 
\end{eqnarray}

where $E(z) = [\Omega_{\rm M} (1+z)^3 + \Omega_{\Delta}]^{-1/2}$. For one line-of-sight,
$\Delta X(z)= X(z_{\rm max}) - X(z_{\rm min})$, with $z_{\rm max}$ being the emission
redshift minus 3000~km~s$^{-1}$ and $z_{\rm min}$ is the redshift of an H~{\sc i} Lyman-$\alpha$
line located at the position of the QSO Lyman-$\beta$ emission line.
In Figure \ref{frequency} we show the function $f(N,X)$ obtained from our statistical sample 
over the redshift range $ z = 2.55 - 5.03$ and for log~$N$(H~{\sc i})~$\geq 19.5$. It can be seen that
there is no break at the low column density end, between log~$N$(H~{\sc i})~=~19.5 and 20.6. 
This makes us confident that we are complete down to log~$N$(H~{\sc i})~=~19.5.
The vertical bars indicate 1$\sigma$ errors. The horizontal bars indicate the bin sizes plotted at the mean column 
density for each bin. PHW05 results in the redshift range $ z = 2.2 - 5.5$ and for log$N$(H~{\sc i})~$\geq 20.3$ 
are overplotted in the same figure. Although our data points are consistent within about 1$\sigma$ with 
those of PHW05, it seems that the overwhole shape of the function is flatter in our data. 
Note that we do not detect any system with log~$N$(H~{\sc i})~$>$~21.25. 
Data points from P\'eroux et al. (2005), hereafter PDDKM05, are also overplotted. Our point at log$N$(H~{\sc i})~=~19.5 is 
consistent with theirs.

A power-law, a gamma function and/or a double 
power-law are usually used to fit the frequency distribution. 
It is apparent from Fig.~\ref{frequency} that a power-law (of the form $f(N,X) = K \times N^{-\alpha}$)
fits the function nicely over the column density range 19.5~$<$~log~$N$(H~{\sc i})~$<$~21. 
The index of this power spectrum (see Table~\ref{par_freq}) is larger ($\alpha$~$\sim$~-1.4) than what is found by 
PHW05 but over a smaller column density range log~$N$(H~{\sc i})~$>$~20.3. 
The discrepancy is apparently due to the difference in the column density ranges
considered by both studies. If we restrict our fit to the same range as PHW05 we find 
an index of $\alpha$~$\sim$~-1.8$\pm0.25$ which is consistent with the results of PHW05.
We note that PDDKM05 already mentioned that the small end of the column density
distribution is flatter than $\alpha$~=~$-2$.

\begin{figure*}[!htb]
\begin{center}
\includegraphics[height=12cm,width=16cm]{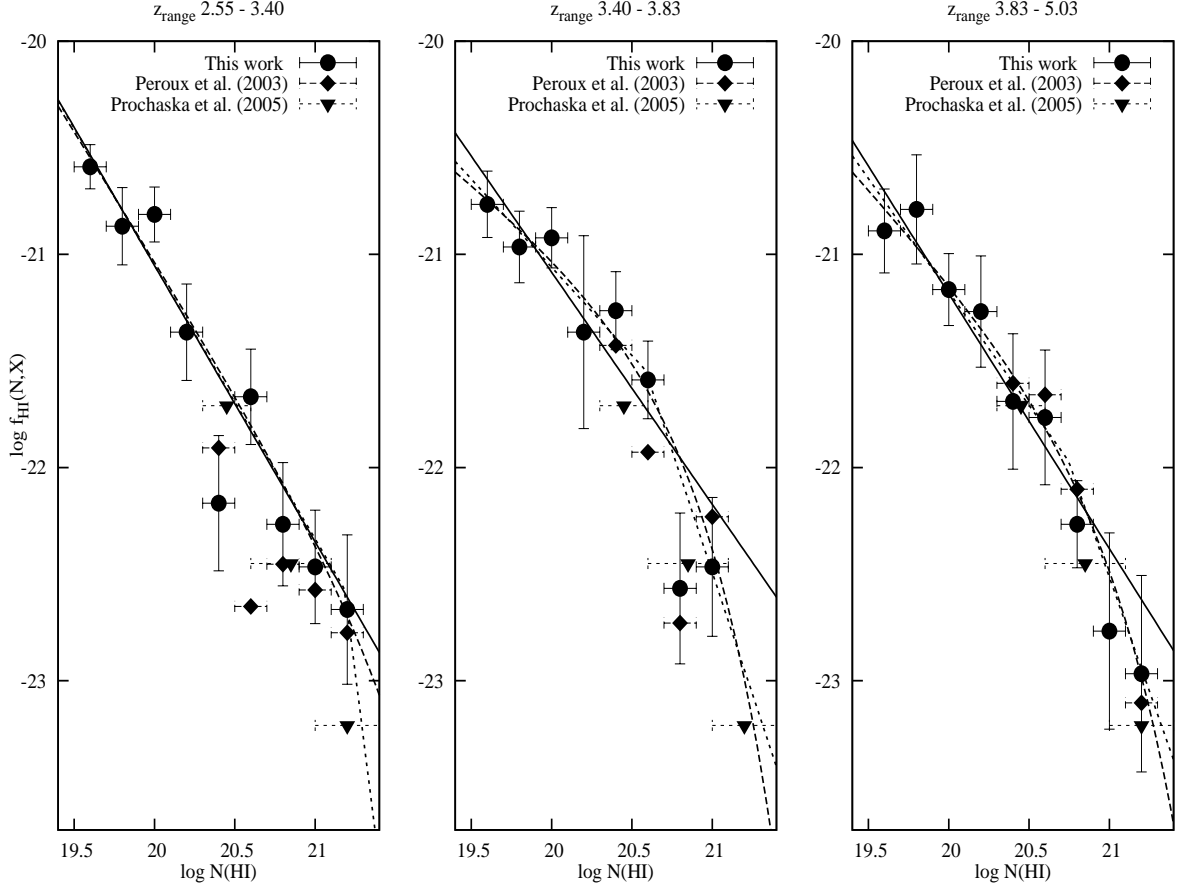}
\caption{H~{\sc i} frequency distribution, $f_{\rm H_I}(N,X)$, 
in three redshifts bins: 2.55$-$3.40 (left), 3.40$-$3.83 (center), and 3.83$-$5.03 (right). 
Straight lines show best $\chi ^2$ fits of a power-law function to the binned data. 
The dashed curve is the same for a gamma function fit. 
and the dot-dashed curve for a double power-law fit.}
\label{painel}
\end{center}
\end{figure*}

There is a large deficit of high column density systems in our survey compared to what
would be expected from the single power-law fit. This has been noted before and discussed
in detail by PHW05. A double power-law was used to fit our full sample with better result
(see Table~\ref{par_freq}). 
However, the sharp break in the function at log~$N$(H~{\sc i})~$\sim$~21 suggests a 
gamma function  of the form, $f(N,X) = K \times (\frac {N} {N_T})^{-\alpha} \times e^{\frac{-N}{N_T}}$ (Pei \& Fall 1995)
should better describe the data. 

We have calculated the frequency distribution function, $f(N,X)$, in different
redshift bins of equal distance path : 2.55$-$3.40; 3.40$-$3.83 and 3.83$-$5.03.
Results are shown in Fig. \ref{painel}. The functions are fitted as described above and
fit results are given in Table \ref{par_freq}. 
We find that the function do not evolve much with redshift.
This is consistent with the finding by PHW05 that the global characteristics of the function are not
changing much with time. There is however a tendancy for a flattening of the function which may indicate
that the number of sub-DLAs relatively to other systems is larger at lower redshift.
Although we do not think this is the case because we have used a conservative
approach, part of this evolution at the highest redshift could possibly be a 
consequence of loosing the sub-DLAs in the strongly blended Lyman-$\alpha$ forest at $z>4$.
We note also a slight decrease of the number of systems with log~$N$(H~{\sc i})~$>$~20.5
at the highest redshifts. This is consistent with the finding by PDDKM05 that
the relative number of high column density DLAs decreases with redshift.
 
Another way to look at these variations is to compute
the redshift evolution of the number of (sub)DLAs per unit path length. 
The observed density of systems is defined as 

\begin{equation}
l_{\rm (sub)DLAs} = \int_{N_{\rm min}}^{N_{\rm max}} f(N,X) dN
\end{equation}
 
\begin{figure}[!htb] 
\begin{center}
\includegraphics[height=9cm,width=9cm]{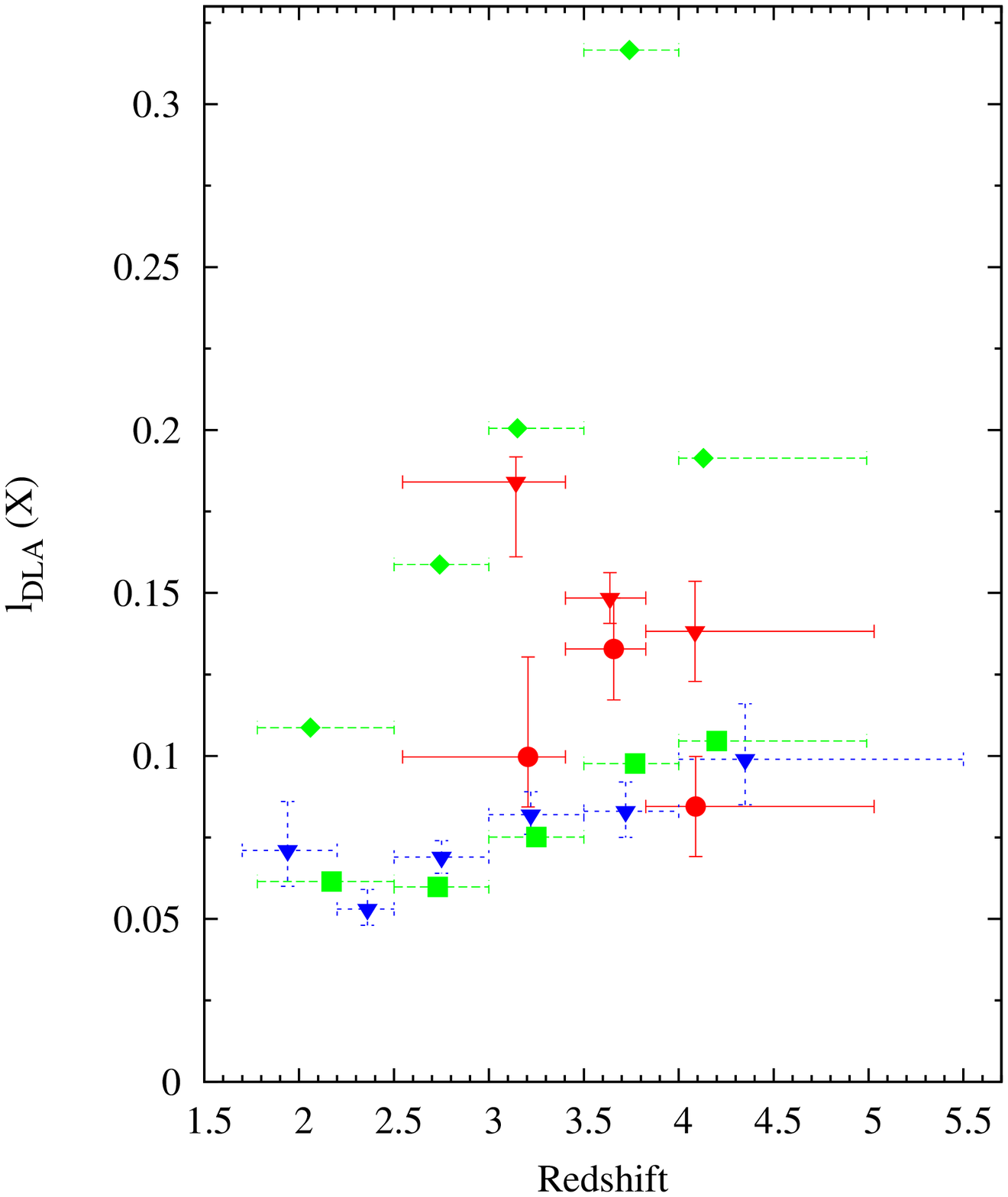}
\caption{Number density of absorbers vs. redshift. Red circles and inverse triangles are from this work
for DLAs and sub-DLAs, respectively. The values obtained by PWH05 (blue inverse triangles) and PDDKM05 (green squares) for DLAs are overplotted. The green diamonds are the values obtained for log N(H~{\sc i}) $>$ 19.0 by PDDKM05.}
\label{ldla}
\end{center}
\end{figure}

Results obtained during the present survey together with those of PWH05 and PDDKM05 are plotted in Figure \ref{ldla}.
The important feature of this plot is that the number density of DLAs peaks at $z\sim 3.5$.
In addition, the ratio of the number of sub-DLAs to the number of DLAs is larger
at redshift $<$3.5 compared to higher redshifts.

\subsection{Neutral hydrogen cosmological mass density, $\Omega_{\rm H_I}$}

\begin{figure}
\begin{center}
\includegraphics[height=9cm,width=9cm]{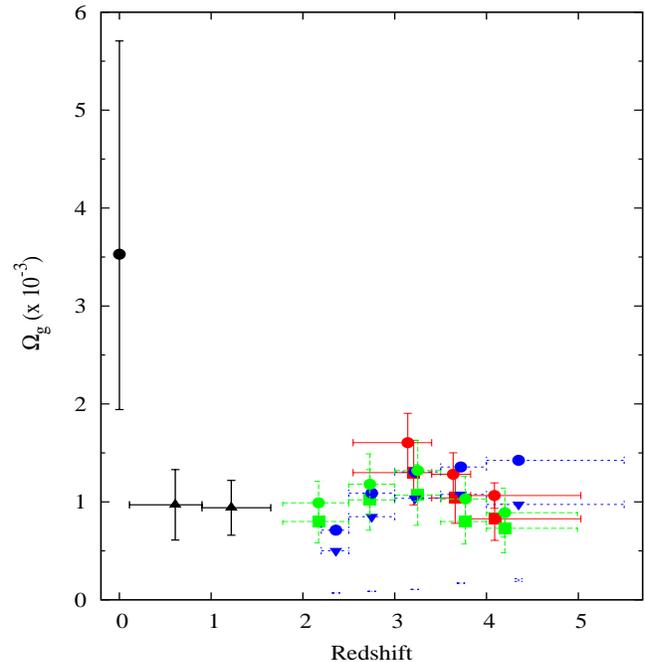}
\caption{Cosmological evolution of the H~{\sc i} mass density. For DLAs: red squares are the results of this work, 
green squares are from PDDKM05, blue inverse triangles from PHW05 and black triangles from Rao et al. (2005). 
For sub-DLAs: red circles are this work (for systems with column densities $\geq$~19.5), 
green circles obtained from PDDKM05 (for systems with column densities $\geq$~19.0) 
and blue circles obtained from PHW05 (as determined from the single power-law fit to the LLS 
frequency distribution function). The black circle are the mass density in stars in local galaxies (Fukugita et al. 1998).}\label{omega_dla}
\end{center}
\end{figure}

The comoving mass density of neutral gas is given by
\begin{equation}
\Omega_{\rm H_I}(z) = \frac{H_0}{c}  \frac{\mu  m_{\rm H}}{\rho_{0}}  \frac{\Sigma_i N_i({\rm H_I})}{\Delta X(z)}
\label{omega}
\end{equation}

where the density is in units of the current critical density $\rho_{0}$; $m_{\rm H}$ is the mass
of the hydrogen atom, $\mu$~=~1.3 is the mean molecular weight, $\Delta X$ the absorption length and the summation
of column densities is done over all absorption systems detected in the survey.
Results are shown in Fig.~\ref{omega_dla} and summarized in Table~\ref{omegaHI}.
The three redshift bins considered are defined so that the absorption length is equal in each bin.
The different columns of Table~\ref{omegaHI} give, respectively, the redshift range, the mean redshift, 
the number of lines-of-sight involved, the number of DLAs and sub-DLAs detected over this redshift range, the
absorption length (calculated using $z_{\rm min}$ and $z_{\rm max}$ as defined in Table~\ref{candidatosdlas}), 
the resulting H~{\sc i} cosmological densities for, respectively, 
DLAs only or both DLAs and sub-DLAs.

\begin{figure}
\begin{center}
\includegraphics[height=9cm,width=9cm]{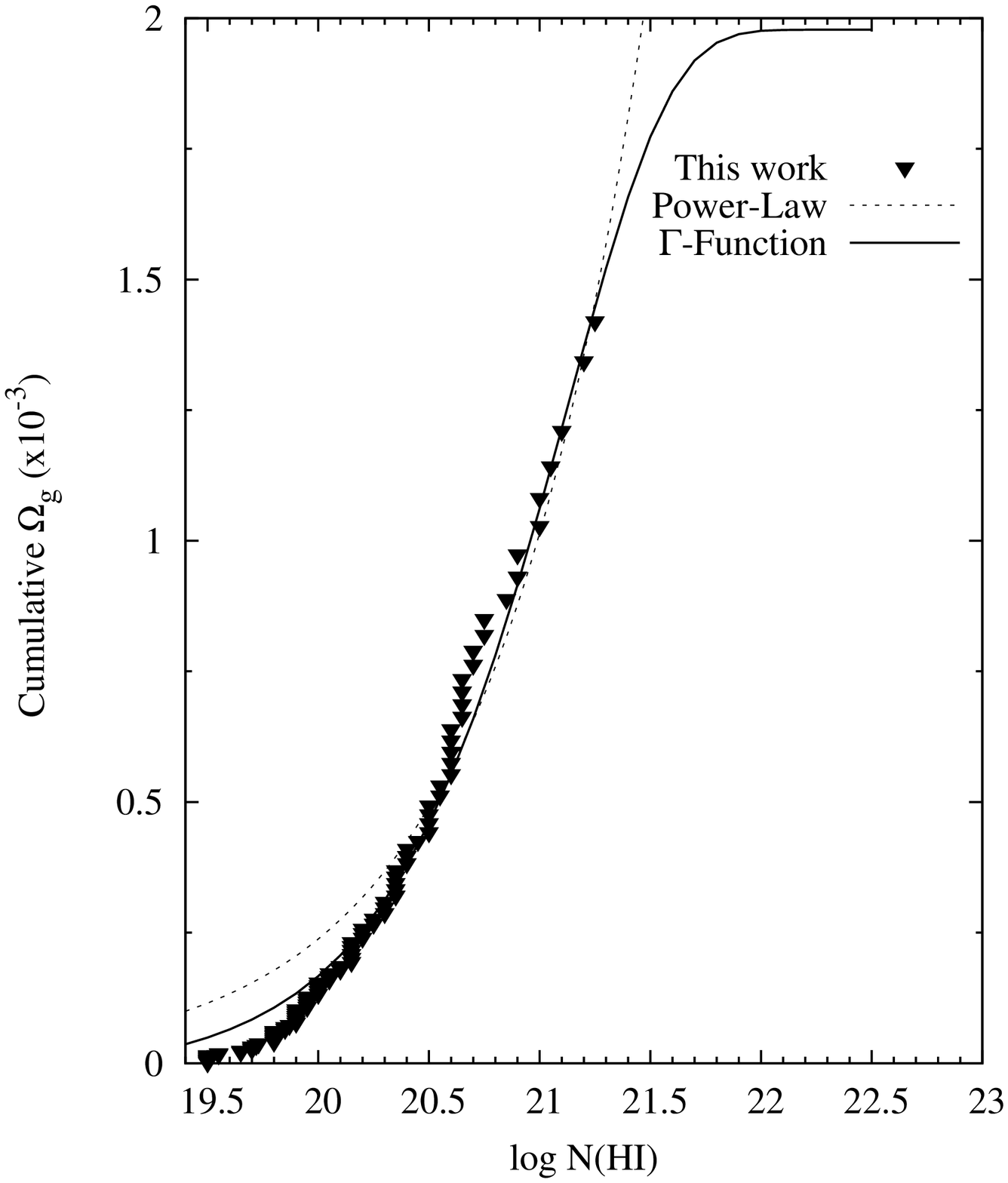}
\caption{$\Omega_{\rm H_I}$ is plotted as a function of the maximum log $N$(H~{\sc i})
considered}\label{Om_NHI}
\end{center}
\end{figure}

Results from PHW05 and PDDKM05 are also plotted on Figure \ref{omega_dla}.
It can be seen that we confirm the decrease of $\Omega_{\rm H_I}$ for $z>3$ that was noticed by
PDDKM05. The measurement from the SDSS in this redshift range is higher.  
However, our survey is of higher spectral resolution and should in principle be more reliable in this redshift range.
It seems that the evolution of $\Omega_{\rm H_I}$ is a 
steep increase from $z=2$ to $z=3$ and then a slightly flatter decrease up to $z=5$. The inclusion of
the sub-DLAs does not change this picture as sub-DLAs contribute to a maximum of about 30\%
to the total H~{\sc i} mass. The contribution by sub-DLAs is better seen in Fig.~\ref{Om_NHI}
where we plot the cumulative density versus the maximum H~{\sc i} column density considered. 
As noted already by numerous authors, the discrepancy of measurements at 
$z<1.5$ is still a problem.

It can be also seen from the Figure \ref{omega_dla} that $\Omega_{\rm H_I}$ are lower than $\Omega_{\rm stellar}$, the 
mass density in stars in local galaxies. We find for the ratio of the peak value of $\Omega_{\rm H_I}$ to 
$\Omega_{\rm stellar}$ for this work R $\sim$ 0.45. Previous surveys, PDDKM05 and PHW05, have found for the 
ratio 0.37 and 0.40 respectively.

\section{Discussion}

We have presented the results of a survey for damped and sub-damped Lyman-$\alpha$ systems 
(log$N$(H~{\sc i})~$\geq$~19.5) at $z \geq 2.55$ along the lines-of-sight to 77 quasars with 
emission redshifts in the range $4 \geq z_{\rm em} \geq 6.3$. In total 99 quasars were observed but 
22 lines-of-sight were not used because of not enough SNR and/or because of
the presence of broad absorption lines.
Intermediate resolution ($R\sim4300$) spectra have been obtained with the 
Echellette Spectrograph and Imager (ESI)  mounted on the Keck telescope. 
The damped Lyman-$\alpha$ absorptions are identified on the basis of
(i) the width of the saturated absorption, (ii) the presence of damped wings and (iii) the presence
of metals at the corresponding redshift. The detection is run automatically but all lines
are verified visually. 
A total of 100 systems with log~$N$(H~{\sc i})~$\geq$~19.5 are detected of which 40 systems
are Damped Lyman-$\alpha$ systems (log~$N$(H~{\sc i})~$\geq$~20.3) for an absorption
length of $\Delta X$~=~378. Spectra are shown in Appendix.

PHW05 have derived from SDSS data that the cosmological density of the neutral gas increases 
strongly by a factor close to two from $z\sim 2$ to $z\sim 3.5$. Beyond this redshift, 
measurements are more difficult because the Lyman-$\alpha$ forest is dense. Our measurements 
should be more reliable because of better spectral resolution.
We show, consistently with the findings of PDDKM05, that the cosmological 
density of the neutral gas decreases at $z>3.5$. 
The overall cosmological evolution seems therefore to have a peak at this redshift. 

We find that the H~{\sc i} column density distribution does not evolve strongly from
$z\sim 2.5$ to $z\sim 4$. The one power-law fit in the range log~$N$(H~{\sc i})~$>$~20.3
gives an index of $\alpha$~=~$-$1.80$\pm$0.25, consistent with previous determinations.
However, we find that the fit over the column density range log~$N$(H~{\sc i})~=~19.5$-$21 
is quite flat ($\alpha$~$\sim$~1.4). This probably indicates that the slope at the low-end
is much flatter than $-2$. This power-law overpredicts data at the high-end
and a second much steeper power-law (or a gamma function) is needed. The fraction
of H~{\sc i} mass in sub-DLAs is of the order of 30\%. Our data do not support
the claim by PDDKM05 that the incidence of low column-density sytems
is larger at high redshift. The number density of sub-DLAs seems to peak at
$z\sim 3.5$ as well.

It is apparent that statistical errors are still large in our survey. It would
be therefore of first importance to enlarge the sample of (sub)DLAs at high redshift.
For this we need to go to fainter quasars however. The advent of X-shooter, a new
generation spectrograph at the VLT with a spectral resolution of $R=6700$ in the 
optical, should allow this to be done in a reasonable amount of observing time.

\begin{acknowledgements}
SGD is supported by the NSF grant AST-0407448,
and the Ajax Foundation.  Cataloguing of DPOSS and discovery of PSS QSOs
was supported by the Norris Foundation and other private donors.
We thank E. Thi\'ebaut, and D. Munro for freely distributing his yorick 
programming language (available at ftp://ftp-icf.llnl.gov:/pub/Yorick), which 
we used to implement our analysis. 
The authors wish to recognize and acknowledge the very significant cutural
role and reverence that the summit of Mauna Kea has always had within the
indigenous Hawaian community. We are most fortunate to have the opportunity
to conduct observations from this mountain. We acknowledge the Keck support
staff for their efforts in performing these observations.
\end{acknowledgements}

\begin{longtable}{l l l l l l l l}
\caption{\label{logbook}Summary of Observations}\\
\hline\hline
QSO           &  $z_{\rm em}$  & $R$ & RA(2000) & DEC(2000) & Obs. Date & Exp. Time (min) &	Note$^*$	\\
\hline
\endfirsthead
\caption{continued.}\\
\hline\hline
QSO           &  $z_{\rm em}$  & $R$ & RA(2000) & DEC(2000) & Obs. Date & Exp. Time (min) &	Note$^*$	\\ 
\hline
\endhead
\hline
\endfoot

PSS0007+2417  & 4.050 & 18.69 & 00 07 38.7  & +24 17 25.0  & 2000 Sep 06 & 60 &  \\ 
PSS0014+3032  & 4.470 & 18.81 & 00 14 42.8  & +30 32 03.0  & 2000 Sep 06 & 60 &  \\
PSS0052+2405  & 4.280 & 18.30 & 00 52 06.8  & +24 05 39.0  & 1999 Dec 30 & 40 &  \\ 
PSS0117+1552  & 4.244 & 18.60 & 01 17 31.2  & +15 52 16.4  & 2000 Sep 04 & 50 & 2\\  
PSS0118+0320  & 4.232 & 18.50 & 01 18 52.8  & +03 20 50.0  & 2000 Sep 06 & 60 & 3\\  
PSS0121+0347  & 4.127 & 17.86 & 01 21 26.2  & +03 47 07.0  & 1999 Dec 30 & 45 & 3\\
SDSS0127-0045 & 4.084 & 18.37 & 01 27 00.7	& -00 45 59.4  & 2001 Jan 02 & 60 &  \\    
PSS0131+0633  & 4.430 & 18.24 & 01 31 12.2  & +06 33 40.0  & 2000 Sep 04 & 60 & 1, 2\\  
PSS0133+0400  & 4.154 & 17.86 & 01 33 40.4  & +04 00 59.0  & 1999 Dec 29 & 40 & 1, 2, 3\\  
PSS0134+3307  & 4.536 & 18.82 & 01 34 21.6  & +33 07 56.5  & 2000 Sep 06 & 60 & 1, 2\\ 
PSS0207+0940  & 4.136 & 18.63 & 02 07 03.5  & +09 40 59.0  & 2000 Sep 04 & 75 &  \\
SDSS0206+1216 & 4.810 & 21.51 & 02 06 51.4  & +12 16 24.4  & 2002 Dec 07 & 30 &  \\
PSS0209+0517  & 4.194 & 17.36 & 02 09 44.7  & +05 17 14.0  & 1999 Dec 29 & 35 & 1, 2, 3\\
SDSS0210-0018 & 4.700 & 20.74 & 02 10 43.2  & -00 18 18.5  & 2002 Dec 07 & 40 & 4, 5\\   
PSS0211+1107  & 3.975 & 18.12 & 02 11 20.1  & +11 07 16.0  & 1999 Dec 29 & 70 &  \\
SDSS0211-0009 & 4.900 & 22.04 & 02 11 02.7  & -00 09 10.3  & 2002 Dec 07 & 20 & 1, 2\\
SDSS0231-0728 & 5.410 & 21.54 & 02 31 37.6  & -07 28 54.5  & 2002 Dec 06 & 90 & 4, 5\\ 
PSS0244-0108  & 3.990 & 19.00 & 02 44 57.2  & -01 08 08.7  & 2000 Sep 06 & 40 & 4, 5\\
PSS0248+1802  & 4.422 & 18.40 & 02 48 54.3  & +18 02 50.3  & 1999 Dec 30 & 60 & 1, 2\\
PSS0320+0208  & 3.960 & 18.74 & 03 20 42.7  & +02 08 16.0  & 1999 Dec 30 & 55 &  \\
SDSS0338+0021 & 5.020 & 21.68 & 03 38 29.3  & +00 21 56.5  & 2002 Dec 06 & 90 & 1, 2, 4, 5\\
SDSS0338-RD657& 4.960 & 23.00 & 03 38 31.3  & +00 18 07.7  & 2001 Jan 02 & 120&  \\
PSS0452+0355  & 4.420 & 18.80 & 04 52 51.5  & +03 55 58.0  & 1999 Dec 30 & 60 &  \\    
PSS0747+4434  & 4.435 & 18.06 & 07 47 50.0  & +44 34 16.0  & 1999 Dec 30 & 90 & 1, 2, 5\\   
SDSS0756+4104 & 5.090 & 21.70 & 07 56 18.0  & +41 04 10.6  & 2001 Mar 26 & 150& 4\\ 
PSS0808+5215  & 4.510 & 18.82 & 08 08 49.5  & +52 15 16.0  & 2000 Apr 28 & 70 & 5\\  
SDSS0810+4603 & 4.074 & 18.67 & 08 10 54.7  & +46 03 55.2  & 2001 Mar 24 & 60 &  \\
PSS0852+5045  & 4.200 & 19.00 & 08 52 27.4  & +50 45 11.0  & 2000 May 13 & 60 &  \\
PSS0926+3055  & 4.190 & 17.31 & 09 26 36.3  & +30 55 06.0  & 1999 Dec 29 & 40 &  \\
SDSS0941+5947 & 4.820 & 20.66 & 09 41 08.4  & +59 47 25.8  & 2002 Dec 06 & 40 & 4, 5\\
BR0945-0411   & 4.130 & 18.80 & 09 47 49.6  & -04 25 15.1  & 2000 May 14 & 60 &  \\   
PSS0950+5801  & 3.973 & 17.38 & 09 50 14.0  & +58 01 38.0  & 1999 Dec 29 & 40 & 4, 5\\ 
BR0951-0450   & 4.350 & 18.90 & 09 53 55.7  & -05 04 19.5  & 2000 May 15 & 53 & 2\\   
PSS0955+5940  & 4.340 & 17.84 & 09 55 11.3  & +59 40 32.0  & 1999 Dec 29 & 56 & 4, 5\\  
PSS0957+3308  & 4.283 & 17.59 & 09 57 44.5  & +33 08 23.0  & 1999 Dec 29 & 55 & 5\\    
BRI1013+0035  & 4.380 & 18.80 & 10 15 49.0  & +00 20 19.0  & 2000 May 15 & 60 & 2\\  
PSS1026+3828  & 4.180 & 18.93 & 10 26 56.7  & +38 28 43.0  & 2000 Apr 29 & 95 & 5\\ 
BR1033-0327   & 4.509 & 18.50 & 10 36 23.7  & -03 43 20.0  & 2000 May 13 & 20 &  \\
PSS1039+3445  & 4.390 & 19.20 & 10 39 19.3  & +34 45 10.9  & 2001 Apr 18 & 150& 5\\
SDSS1044-0125 & 5.740 & 25.10 & 10 44 33.0  & -01 25 03.1  & 2000 Apr 28 \& 2001 Jan 01,02 & 330 & \\
SDSS1048+4637 & 6.230 & 22.40 & 10 48 45.0  & +46 37 18.3  & 2003 Jun 02 & 90 & 5\\
PSS1048+4407  & 4.450 & 19.50 & 10 48 46.6  & +44 07 12.7  & 2000 Apr 29 & 20 &  \\    
PSS1057+4555  & 4.126 & 17.70 & 10 57 56.4  & +45 55 51.9  & 1999 Dec 30 & 40 & 1, 2, 5\\   
PSS1058+1245  & 4.330 & 18.00 & 10 58 58.5  & +12 45 55.0  & 1999 Dec 30 & 75 &  \\  
PSS1118+3702  & 4.030 & 18.76 & 11 18 56.2  & +37 02 53.9  & 2001 Mar 24 & 60 & 5\\ 
PSS1140+6205  & 4.509 & 18.73 & 11 40 09.6  & +62 05 23.3  & 2000 May 14 & 60 & 4, 5\\ 
PSS1159+1337  & 4.081 & 18.50 & 11 59 06.5  & +13 37 37.8  & 2000 Feb 10 & 55 & 1, 2, 5\\
SDSS1204-0021 & 5.030 & 20.82 & 12 04 41.7  & -00 21 49.6  & 2003 Jun 04 & 60 & 4, 5\\
SDSS1208+0010 & 5.273 & 22.75 & 12 08 23.9  & +00 10 28.9  & 2001 Mar 24 & 120&  \\
PSS1226+0950  & 4.340 & 18.78 & 12 26 23.8  & +09 50 03.7  & 2001 Mar 24 & 80 &  \\
PSS1247+3406  & 4.897 & 20.40 & 12 49 42.2  & +33 49 54.0  & 2001 Mar 26 & 60 &  \\
PSS1248+3110  & 4.346 & 18.90 & 12 48 20.2  & +31 10 44.0  & 2000 May 13 & 60 &  \\   
PSS1253-0228  & 4.007 & 19.40 & 12 53 36.3  & -02 28 08.0  & 2000 Apr 29 & 55 & 1,2\\ 
SDSS1310-0055 & 4.151 & 18.85 & 13 10 52.6  & -00 55 31.8  & 2001 Mar 24 & 50 & 4, 5\\
PSS1315+2924  & 4.180 & 18.48 & 13 15 39.6  & +29 24 39.8  & 2001 Mar 24 & 85 &  \\ 
PSS1317+3531  & 4.369 & 19.10 & 13 17 43.3  & +35 31 33.1  & 2001 Mar 26 & 55 & 2\\
J1325+1123    & 4.400 & 18.77 & 13 25 12.6  & +11 23 32.8  & 2001 Apr 18 & 120& 5\\    
PSS1326+0743  & 4.123 & 17.30 & 13 26 11.9  & +07 43 59.0  & 2000 Dec 29 & 60 &  \\
PSS1339+5154  & 4.080 & 18.70 & 13 39 13.0  & +51 54 04.0  & 2000 May 15 & 60 & 5\\
PSS1347+4956  & 4.560 & 17.90 & 13 47 43.4  & +49 56 21.0  & 2000 Feb 10 & 50 & 5\\
PSS1401+4111  & 4.026 & 18.62 & 14 01 32.8  & +41 11 49.9  & 2000 Apr 28 & 46 & 4, 5\\
PSS1403+4126  & 3.862 & 18.92 & 14 03 55.7  & +41 26 16.2  & 2000 May 15 & 90 & 4, 5\\
PSS1418+4449  & 4.280 & 18.40 & 14 18 31.8  & +44 49 37.0  & 2000 Feb 29 & 60 & 4, 5\\
PSS1430+2828  & 4.306 & 19.30 & 14 30 31.9  & +28 28 34.1  & 2001 Mar 26 & 55 & 2\\
PSS1432+3940  & 4.292 & 18.60 & 14 32 24.9  & +39 40 24.0  & 2000 May 13 & 60 & 4, 5\\     
PSS1435+3057  & 4.350 & 19.30 & 14 35 23.3  & +30 57 16.3  & 2001 Apr 18 & 70 & 2\\   
PSS1443+2724  & 4.406 & 19.30 & 14 43 31.2  & +27 24 37.0  & 2000 Apr 29 & 30 & 2\\
PSS1443+5856  & 4.270 & 17.80 & 14 43 40.8  & +58 56 53.0  & 2000 Apr 28 & 40 & 4, 5\\
PSS1458+6813  & 4.291 & 18.67 & 14 58 31.7 	& +68 13 05.2  & 2000 Apr 28 & 60 &  \\  
PSS1500+5829  & 4.224 & 18.60 & 15 00 07.7  & +58 29 38.0  & 2000 May 14 & 60 &  \\  
PSS1506+5220  & 4.180 & 18.10 & 15 06 54.6  & +52 20 05.0  & 2000 Feb 10 & 50 & 4, 5\\
GB1508+5714   & 4.304 & 18.90 & 15 10 02.2  & +57 03 04.9  & 2000 May 13 & 75 & 2\\  
PSS1531+4157  & 4.200 & 19.00 & 15 31 29.4  & +45 17 07.9  & 2001 Apr 18 & 128&  \\  
PSS1535+2943  & 3.972 & 18.90 & 15 35 53.9  & +29 43 13.0  & 2000 May 15 \& 2000 Sep 05 & 120 & \\
PSS1543+3417  & 4.390 & 18.40 & 15 43 40.4  & +34 17 45.0  & 2000 Apr 30 & 76 & 5\\                 
PSS1554+1835  & 3.990 & 18.90 & 15 35 53.9  & +29 43 13.0  & 2000 May 15 \& 2000 Sep 04 & 120 &\\   
PSS1555+2003  & 4.228 & 18.90 & 15 55 02.6  & +20 03 25.0  & 2000 May 14 & 90 &  \\ 
PSS1615+1803  & 4.010 & 18.42 & 16 15 22.9  & +18 03 56.4  & 2000 Apr 29 & 90 &  \\
SDSS1630+4012 & 6.050 & 20.42 & 16 30 33.9  & +40 12 09.6  & 2003 Jun 02,03,04 &  295 & \\
PSS1633+1411  & 4.349 & 19.00 & 16 33 19.7  & +14 11 42.6  & 2000 Apr 30 & 60 & 1, 2\\  
PSS1646+5514  & 4.084 & 18.11 & 16 46 56.3  & +55 14 46.7  & 2000 Apr 28 & 60 & 1, 2\\
VLA1713+4218  & 4.230 & 19.00 & 17 13 56.2  & +42 18 08.6  & 2001 Apr 17 & 90 &  \\   
PSS1715+3809  & 4.520 & 18.56 & 17 15 39.5  & +38 09 06.6  & 2000 Sep 03 & 60 &  \\  
PSS1721+3256  & 4.040 & 19.23 & 17 21 06.7  & +32 56 35.8  & 2000 May 01 & 60 & 1, 2\\
PSS1723+2243  & 4.514 & 18.17 & 17 23 23.1  & +22 43 56.4  & 2001 Mar 24 & 40 &  \\
SDSS1737+5828 & 4.940 & 20.93 & 17 37 44.0  & +58 28 25.4  & 2001 Mar 24 & 90 & 4, 5\\  
PSS1745+6846  & 4.130 & 19.12 & 17 45 50.1  & +68 46 21.0  & 2000 May 01 & 120 & \\  
PSS1802+5616  & 4.158 & 19.19 & 18 02 48.9  & +56 16 51.0  & 2000 Sep 05 & 90 & 1, 2\\   
PSS2122-0014  & 4.114 & 19.13 & 21 22 07.5  & -00 14 45.0  & 2000 Sep 06 & 60 & 1, 2, 4, 5\\   
PSS2154+0335  & 4.360 & 18.41 & 21 54 06.9  & +03 35 40.0  & 2000 May 14 & 60 & 1, 2, 3\\
PSS2155+1358  & 4.256 & 18.50 & 21 55 02.1  & +13 58 26.0  & 2000 May 15 & 40 & 1, 2, 3 \\   
PSS2203+1824  & 4.375 & 18.74 & 22 03 43.4  & +18 28 14.0  & 2000 Sep 04 & 60 &  \\  
PSS2238+2603  & 4.031 & 18.85 & 22 38 41.1  & +26 03 46.0  & 2000 Sep 06 & 60 &  \\  
PSS2241+1352  & 4.441 & 18.69 & 22 41 47.7  & +13 52 05.9  & 2000 Sep 03 & 60 & 1, 2, 4, 5\\
PSS2244+1005  & 4.040 & 18.92 & 22 44 05.5  & +10 47 38.0  & 2000 Sep 04 & 80 &  \\  
PSS2315+0921  & 4.412 & 18.96 & 23 15 59.2  & +09 21 44.0  & 2000 Sep 05 & 90 &  \\  
PSS2322+1944  & 4.170 & 18.29 & 23 22 07.2  & +19 44 23.0  & 2000 May 13 \& 2000 Sep 04 & 111 & \\
PSS2323+2758  & 4.180 & 18.51 & 23 23 41.0  & +27 58 01.0  & 2000 Sep 04 & 86 &  \\  
PSS2344+0342  & 4.340 & 17.87 & 23 44 03.2  & +03 42 26.0  & 1999 Dec 30 & 45 & 1, 2 \\  
\hline\\
\multicolumn{8}{l}{$^*$ Quasars from this survey were previously used for $\Omega_{\rm DLA, HI}$ estimations by: (1) P\'eroux et al. (2001);}\\
\multicolumn{8}{l}{(2) P\'eroux et al. (2003); (3) P\'eroux et al. (2005); (4) Prochaska et al. (2005); (5) Prochaska \& Wolfe (2009)}
\end{longtable}

\begin{longtable}{l l l l l l l l}
\caption{\label{candidatosdlas}List of DLA-subDLAs}\\
\hline\hline
QSO           &  $z_{\rm em}$  & $z_{\rm min}$& $z_{\rm max}$ & \multicolumn{2}{c}{DLA-subDLA}             & Metal Lines  & Note~$^*$   						\\       
              &                & 	          &               & $z_{\rm abs}$      &   log $N$(H~{\sc i})  &              &     								\\
\hline
\endfirsthead
\caption{continued.}\\
\hline\hline
QSO           &  $z_{\rm em}$  & $z_{\rm min}$& $z_{\rm max}$ & \multicolumn{2}{c}{DLA-subDLA}            & Metal Lines           & Note~$^*$    				\\       
              &                & 	          &               & $z_{\rm abs}$     &   log $N$(H~{\sc i})  &     				  &				                \\
\hline
\endhead
\hline
\endfoot
PSS0007+2417  & 4.05  & 2.758 & 4.000 & 3.194 & 19.50$\pm$0.07 & Si~{\sc ii}, C~{\sc iv}										    				    &\\  
              &       &       &       & 3.497 & 20.95$\pm$0.10 & Si~{\sc ii}, Fe~{\sc ii}, Al~{\sc ii} 								                    &\\  
              &       &       &       & 3.706 & 20.65$\pm$0.15 & Si~{\sc ii}, O~{\sc i}, C~{\sc ii}, Si~{\sc iv}, C~{\sc iv}, Fe~{\sc ii}, Al~{\sc ii}  &\\  
              &       &       &       & 3.839 & 20.65$\pm$0.10 & Si~{\sc ii}, O~{\sc i}, C~{\sc ii}, C~{\sc iv}, Fe~{\sc ii}, Al~{\sc ii}               &\\  
PSS0117+1552  & 4.244 & 2.784 & 4.192 & 3.175 & 19.50$\pm$0.10 & C~{\sc iv}																				&\\  
PSS0118+0320  & 4.232 & 2.901 & 4.180 & 4.128 & 19.95$\pm$0.15 & Si~{\sc ii}, O~{\sc i}, C~{\sc ii}, Si~{\sc iv}, C~{\sc iv}, Al~{\sc ii}  				& 3\\  
PSS0121+0347  & 4.127 & 2.789 & 4.076 & 2.977 & 19.50$\pm$0.15 & no metals~$^{a}$ 																		& 3\\
SDSS0127-0045 & 4.084 & 2.546 & 4.033 & 3.728 & 20.85$\pm$0.10 & Si~{\sc ii}, O~{\sc i}, C~{\sc ii}, Si~{\sc iv}, C~{\sc iv}, Fe~{\sc ii}, Al~{\sc ii}  &\\    
PSS0131+0633  & 4.430 & 3.008 & 4.376 & 3.173 & 19.95$\pm$0.10 & Fe~{\sc ii}, Al~{\sc ii}																&1, 2\\  
	          &       &       &       & 3.689 & 19.50$\pm$0.10 & C~{\sc iv}																				&1, 2\\  
PSS0133+0400  & 4.154 & 2.930 & 4.102 & 3.690 & 20.40$\pm$0.15 & Si~{\sc ii}																			&1, 2, 3\\  
              &       &       &       & 3.771 & 20.45$\pm$0.15 & Si~{\sc ii}, Si~{\sc iv}, C~{\sc iv}, Fe~{\sc ii}, Al~{\sc ii}							&1, 2, 3\\  
              &       &       &       & 3.995 & 20.15$\pm$0.10 & O~{\sc i}, C~{\sc ii}, Si~{\sc iv}, C~{\sc iv}, Fe~{\sc ii}							&3\\  
PSS0134+3307  & 4.536 & 2.976 & 4.481 & 3.760 & 20.60$\pm$0.10 & Si~{\sc ii}, Si~{\sc iv}, C~{\sc iv}, Al~{\sc ii}										&1, 2\\  
PSS0209+0517  & 4.194 & 2.943 & 4.142 & 3.666 & 20.50$\pm$0.10 & Si~{\sc ii}, Fe~{\sc ii}, Al~{\sc ii}													&1, 2, 3\\  
              &       &       &       & 3.862 & 20.30$\pm$0.15 & O~{\sc i}, C~{\sc ii}, Si~{\sc ii}, Al~{\sc ii}										&1, 2, 3\\  
PSS0211+1107  & 3.975 & 2.847 & 3.925 & 3.139 & 20.05$\pm$0.25 & Si~{\sc ii}, C~{\sc iv}, Fe~{\sc ii}, Al~{\sc ii}										&\\  
              &       &       &       & 3.502 & 20.10$\pm$0.20 & Si~{\sc ii}, C~{\sc iv}, Fe~{\sc ii}, Al~{\sc ii}										&\\  
PSS0747+4434  & 4.435 & 3.107 & 4.381 & 3.762 & 20.00$\pm$0.20 & no metals~$^{d}$																	    &1, 2\\  
              &       &       &       & 4.019 & 21.10$\pm$0.15 & O~{\sc i}, C~{\sc ii}, Si~{\sc ii}, Fe~{\sc ii}, Al~{\sc ii}						    &1, 2, 5\\ 
SDSS0756+4104 & 5.09  & 3.551 & 5.029 & 4.360 & 20.15$\pm$0.10 & no metals																			    &\\ 
PSS0808+5215  & 4.510 & 3.181 & 4.455 & 3.114 & 20.60$\pm$0.10 & no metals~$^{b}$																	    &5\\  
SDSS0810+4603 & 4.074 & 2.811 & 4.023 & 2.956 & 20.55$\pm$0.25 & Al~{\sc iii}																		    &\\  
	          &       &       &       & 3.472 & 19.90$\pm$0.07 & Fe~{\sc ii}, Al~{\sc ii}															    &\\   
PSS0950+5801  & 3.973 & 2.759 & 3.923 & 3.264 & 20.50$\pm$0.10 & Si~{\sc iii} , Si~{\sc ii}, C~{\sc iv}, Fe~{\sc ii}, Al~{\sc ii}					    &4, 5\\ 
BR0951-0450   & 4.35  & 2.866 & 4.296 & 3.235 & 20.25$\pm$0.10 & C~{\sc iv}, Fe~{\sc ii}, Al~{\sc ii}												 	&\\  
	          &       &       &       & 3.856 & 20.70$\pm$0.20 & Si~{\sc ii}, Si~{\sc iv}, C~{\sc iv}, Al~{\sc ii}									    &2\\  
              &       &       &       & 4.202 & 20.35$\pm$0.15 & Si~{\sc ii}, O~{\sc i}, C~{\sc iv}													    &2\\  
PSS0955+5940  & 4.34  & 3.070 & 4.287 & 3.542 & 20.30$\pm$0.10 & Si~{\sc ii}, C~{\sc iv}, Fe~{\sc ii}, Al~{\sc ii}									    &4, 5\\  
              &       &       &       & 3.843 & 20.00$\pm$0.15 & Si~{\sc iv}, Si~{\sc ii}, C~{\sc iv}, Al~{\sc ii}									    &4, 5\\  
              &       &       &       & 4.044 & 20.10$\pm$0.15 & O~{\sc i}, C~{\sc ii}, Si~{\sc ii}, Si~{\sc iv}, C~{\sc iv}						    &\\  
PSS0957+3308  & 4.283 & 2.969 & 4.230 & 3.043 & 19.65$\pm$0.15 & C~{\sc ii}, Al~{\sc ii}															    &\\   
              &       &       &       & 3.280 & 20.50$\pm$0.10 & Fe~{\sc ii}, Al~{\sc ii}															    &\\ 
              &       &       &       & 3.364 & 19.70$\pm$0.15 & Al~{\sc ii}																		    &\\  
              &       &       &       & 3.900 & 19.50$\pm$0.10 & no metals																			    &\\  
              &       &       &       & 4.179 & 20.60$\pm$0.10 & O~{\sc i}, C~{\sc ii}, Si~{\sc ii}, Si~{\sc iv}, C~{\sc iv}						    &\\ 
BRI1013+0035  & 4.38  & 3.123 & 4.326 & 3.738 & 20.15$\pm$0.20 & no metals																			    &\\  
PSS1026+3828  & 4.18  & 2.756 & 4.128 & 3.339 & 19.72$\pm$0.08 & no metals over available coverage~$^c$												    &\\  
              &       &       &       & 3.865 & 19.90$\pm$0.10 & C~{\sc ii}																			    &\\  
PSS1039+3445  & 4.39  & 2.881 & 4.336 & 4.098 & 19.65$\pm$0.10 & O~{\sc i}, C~{\sc ii}, Si~{\sc iV}														&\\  
PSS1057+4555  & 4.126 & 2.776 & 4.075 & 2.909 & 20.05$\pm$0.10 & Fe~{\sc ii}, Al~{\sc ii}																&1, 2\\  
  	          &       &       &       & 3.058 & 19.80$\pm$0.15 & C~{\sc iV}, Fe~{\sc ii}, Al~{\sc ii}													&1, 2\\  
              &       &       &       & 3.164 & 19.50$\pm$0.20 & Si~{\sc ii}, C~{\sc iV}, Fe~{\sc ii}, Al~{\sc ii}										&\\  
              &       &       &       & 3.317 & 20.15$\pm$0.10 & Si~{\sc ii}, Fe~{\sc ii}, Al~{\sc ii}													&1, 2\\  
PSS1058+1245  & 4.330 & 3.110 & 4.277 & 3.430 & 20.40$\pm$0.15 & no metals over available coverage~$^c$													&\\  
PSS1118+3702  & 4.03  & 2.893 & 3.980 & 3.698 & 19.87$\pm$0.08 & O~{\sc i}, C~{\sc ii}, Si~{\sc ii}, Al~{\sc ii}										&\\  
PSS1159+1337  & 4.081 & 2.823 & 4.030 & 3.726 & 20.00$\pm$0.10 & C~{\sc ii}, Si~{\sc ii}, Si~{\sc iV}, C~{\sc iV}										&1, 2\\
PSS1247+3406  & 4.897 & 3.314 & 4.838 & 4.572 & 19.80$\pm$0.10 & Si~{\sc iV}, C~{\sc iV}																&\\
PSS1248+3110  & 4.346 & 3.088 & 4.293 & 3.697 & 20.35$\pm$0.10 & Si~{\sc ii}, Fe~{\sc ii}, Al~{\sc ii}													&\\  
              &       &       &       & 4.075 & 20.20$\pm$0.15 & O~{\sc i}, C~{\sc ii}, Si~{\sc iv}													 	&\\ 
PSS1253-0228  & 4.007 & 2.868 & 3.957 & 3.608 & 19.95$\pm$0.10 & C~{\sc ii}, Si~{\sc iv}, C~{\sc iV}, Fe~{\sc ii}										&1, 2\\  
PSS1317+3531  & 4.369 & 2.998 & 4.315 & 3.461 & 19.90$\pm$0.05 & Si~{\sc ii}, C~{\sc iV}, Fe~{\sc ii}, Al~{\sc ii}										&\\
J1325+1123    & 4.400 & 3.017 & 4.346 & 3.723 & 19.50$\pm$0.20 & C~{\sc iV}																				&\\  
              &       &       &       & 4.133 & 19.50$\pm$0.20 & no metals~$^b$																		    &\\   
PSS1326+0743  & 4.123 & 2.866 & 4.072 & 2.919 & 19.95$\pm$0.10 & Al~{\sc ii}																			&\\  
              &       &       &       & 3.425 & 19.90$\pm$0.15 & Al~{\sc ii}																			&\\  
PSS1432+3940  & 4.292 & 3.105 & 4.239 & 3.274 & 21.00$\pm$0.15 & C~{\sc iv}, Fe~{\sc ii}, Al~{\sc ii}													&\\  
PSS1435+3057  & 4.35  & 2.942 & 4.297 & 3.267 & 20.05$\pm$0.10 & C~{\sc iv}, Al~{\sc ii}																&\\  
              &       &       &       & 3.516 & 20.20$\pm$0.10 & Si~{\sc ii}, C~{\sc iv}, Al~{\sc ii}													&\\  
              &       &       &       & 3.778 & 19.85$\pm$0.10 & C~{\sc iv}, Al~{\sc ii}																&\\  
PSS1443+2724  & 4.406 & 3.095 & 4.352 & 4.223 & 20.90$\pm$0.15 & Si~{\sc iv}, Si~{\sc ii}, C~{\sc iv}, Fe~{\sc ii}, Al~{\sc ii}							&2\\  
PSS1500+5829  & 4.224 & 3.031 & 4.172 & 3.585 & 19.55$\pm$0.10 & no metals over available coverage~$^c$													&\\  
              &       &       &       & 3.915 & 20.50$\pm$0.10 & O~{\sc i}																				&\\  
PSS1506+5220  & 4.18  & 3.275 & 4.128 & 3.223 & 20.70$\pm$0.10 & Si~{\sc ii}, C~{\sc iv}, Fe~{\sc ii}, Al~{\sc ii}, Al~{\sc iii}  						&\\  
PSS1531+4157  & 4.20  & 2.868 & 4.148 & 3.657 & 19.55$\pm$0.15 & Si~{\sc ii}, C~{\sc iv}, Al~{\sc ii}													&\\  
PSS1535+2943  & 3.972 & 2.797 & 3.922 & 3.202 & 20.60$\pm$0.10 & Si~{\sc ii}, C~{\sc iv}, Fe~{\sc ii}, Al~{\sc ii}										&\\  
              &       &       &       & 3.762 & 20.55$\pm$0.15 & O~{\sc i}, C~{\sc ii}, Si~{\sc iv}, C~{\sc iv}, Al~{\sc ii}							&\\  
PSS1554+1835  & 3.99  & 2.769 & 3.940 & 2.919 & 20.00$\pm$0.15 & Si~{\sc ii}, C~{\sc iv}, Al~{\sc ii}, Al~{\sc iii}										&\\  
PSS1555+2003  & 4.228 & 2.824 & 4.176 & 3.427 & 19.90$\pm$0.10 & no metals over available coverage~$^c$													&\\  
PSS1633+1411  & 4.349 & 2.836 & 4.296 & 2.880 & 20.30$\pm$0.15 & Fe~{\sc ii}, Si~{\sc ii}																&\\  
PSS1646+5514  & 4.084 & 2.800 & 4.033 & 2.932 & 19.50$\pm$0.10 & no metals over available coverage~$^c$													&\\  
              &       &       &       & 4.029 & 19.80$\pm$0.15 & no metals over available coverage~$^c$													&\\  
PSS1715+3809  & 4.52  & 3.004 & 4.465 & 3.341 & 21.05$\pm$0.20 & Si~{\sc ii}, Fe~{\sc ii}, Al~{\sc ii}													&\\  
PSS1723+2243  & 4.515 & 3.062 & 4.460 & 3.697 & 20.35$\pm$0.10 & Si~{\sc ii}, C~{\sc iv}, Fe~{\sc ii}, Al~{\sc ii}										&\\
SDSS1737+5828 & 4.94  & 3.383 & 4.881 & 4.152 & 19.85$\pm$0.15 & Si~{\sc ii}, Al~{\sc ii}																&\\  
              &       &       &       & 4.741 & 20.65$\pm$0.15 & Si~{\sc ii}, O~{\sc i}, C~{\sc ii}														&4, 5\\  
PSS1745+6846  & 4.13  & 2.860 & 4.079 & 3.706 & 19.73$\pm$0.09 & Si~{\sc ii}, C~{\sc iv}, Al~{\sc iii}													&\\  
PSS1802+5616  & 4.158 & 2.821 & 4.106 & 3.391 & 20.25$\pm$0.10 & Si~{\sc iv}, Si~{\sc ii}, C~{\sc iv}, Fe~{\sc ii}, Al~{\sc ii} 						&1, 2\\  
              &       &       &       & 3.554 & 20.40$\pm$0.10 & O~{\sc i}, Si~{\sc ii}, C~{\sc iv}, Fe~{\sc ii}, Al~{\sc ii}   						&1, 2\\  
              &       &       &       & 3.762 & 20.65$\pm$0.15 & C~{\sc ii}, Si~{\sc ii}, Fe~{\sc ii}, Al~{\sc ii}										&1, 2\\  
              &       &       &       & 3.811 & 20.35$\pm$0.15 & C~{\sc ii}, Si~{\sc iv}, Si~{\sc ii}, Al~{\sc ii}										&1, 2\\  
PSS2122-0014  & 4.114 & 2.903 & 4.063 & 3.207 & 20.20$\pm$0.10 & Si~{\sc ii}, C~{\sc iv}, Fe~{\sc ii}, Al~{\sc ii}										&1, 2, 4, 5\\  
              &       &       &       & 3.264 & 19.90$\pm$0.15 & C~{\sc iv}, Al~{\sc ii} 																&\\  
              &       &       &       & 4.001 & 20.15$\pm$0.15 & C~{\sc ii}, Si~{\sc iv}, C~{\sc iv}													&1, 2\\  
PSS2155+1358  & 4.256 & 3.064 & 4.203 & 3.143 & 19.90$\pm$0.20 & Fe~{\sc ii}, Al~{\sc ii}																&\\  
              &       &       &       & 3.318 & 20.75$\pm$0.20 & Si~{\sc ii}, Fe~{\sc ii}, Al~{\sc ii}													&1, 2\\  
              &       &       &       & 4.211 & 20.00$\pm$0.10 & Si~{\sc ii}, O~{\sc i}, C~{\sc ii}														&\\  
PSS2203+1824  & 4.375 & 2.850 & 4.321 & 3.610 & 19.70$\pm$0.15 & Si~{\sc ii}, C~{\sc iv}, Al~{\sc ii}													&\\  
              &       &       &       & 4.107 & 19.80$\pm$0.15 & C~{\sc ii}, Si~{\sc ii}, Fe~{\sc ii}, Al~{\sc ii}										&\\  
PSS2238+2603  & 4.031 & 2.816 & 3.981 & 3.857 & 20.45$\pm$0.15 & Si~{\sc ii}, O~{\sc i}, C~{\sc ii}, Si~{\sc iv}, C~{\sc iv}, Fe~{\sc ii}, Al~{\sc ii}	&\\  
PSS2241+1352  & 4.441 & 3.106 & 4.387 & 3.656 & 20.15$\pm$0.20 & Al~{\sc ii}, Si~{\sc ii}																&1, 2\\  
              &       &       &       & 4.281 & 20.75$\pm$0.15 & Si~{\sc ii}, O~{\sc i}, C~{\sc ii}, Fe~{\sc ii}, Al~{\sc ii}							&1, 2, 4, 5\\  
PSS2315+0921  & 4.412 & 2.863 & 4.358 & 3.220 & 21.25$\pm$0.15 & C~{\sc ii}, Si~{\sc ii}, Fe~{\sc ii}, Al~{\sc ii}										&\\  
              &       &       &       & 3.425 & 21.00$\pm$0.20 & Si~{\sc ii}, C~{\sc iv}, Fe~{\sc ii}, Al~{\sc ii}										&\\  
PSS2322+1944  & 4.17  & 2.754 & 4.118 & 2.888 & 19.95$\pm$0.10 & Fe~{\sc ii}, Al~{\sc ii}																&\\  
              &       &       &       & 2.975 & 19.80$\pm$0.10 & Fe~{\sc ii}, Al~{\sc ii}																&\\  
PSS2323+2758  & 4.18  & 2.823 & 4.128 & 2.952 & 19.80$\pm$0.10 & no metals 																				&\\  
              &       &       &       & 3.684 & 20.60$\pm$0.15 & Si~{\sc ii}, Fe~{\sc ii}																&\\  
PSS2344+0342  & 4.340 & 2.939 & 4.287 & 3.220 & 21.20$\pm$0.10 & no metals over available coverage~$^{c, d}$											&1, 2\\  
              &       &       &       & 3.884 & 19.80$\pm$0.10 & no metals over available coverage~$^{c, e}$											&\\
\hline
\multicolumn{6}{l}{$^a$ Metals were detected by P\'eroux et al. (2005)}\\
\multicolumn{6}{l}{$^b$ Metals were detected by Prochaska \& Wolfe (2009)}\\
\multicolumn{6}{l}{$^c$ We do not cover the red part of the spectrum $\lambda \geq$ 6400~$\AA$}\\
\multicolumn{6}{l}{$^d$ Metals were detected by P\'eroux et al. (2001)}\\
\multicolumn{6}{l}{$^e$ Metals were detected by Dessauges-Zavadsky et al. (2003)}\\
\multicolumn{8}{l}{$^*$ The same references that appears in Table \ref{logbook}}
\end{longtable}

\begin{table}[!htb]
\begin{center}
\footnotesize \small
\caption{QSOs without detected DLAs and/or sub-DLAs}
\label{wbas} \vspace{1pc}
\begin{tabular}{l l l l l}
\hline
\hline
QSO           & $z_{em}$ & $z_{min}$ & $z_{max}$  & Note$^*$	\\
\hline
\hline  
PSS0014+3032  & 4.470 & 2.866 & 4.415  & 	\\
SDSS0210-0018 & 4.700 & 3.177 & 4.643  & 4, 5	\\  
PSS0248+1802  & 4.430 & 3.118 & 4.376  & 2	\\  
PSS0452+0355  & 4.395 & 3.115 & 4.341  & 	\\  
PSS0852+5045  & 4.216 & 2.939 & 4.164  &	\\  
PSS0926+3055  & 4.198 & 2.951 & 4.146  &	\\
SDSS0941+5947 & 4.820 & 3.322 & 4.762  & 4, 5	\\    
PSS1140+6205  & 4.509 & 3.113 & 4.454  & 4, 5	\\
SDSS1310-0055 & 4.152 & 2.830 & 4.101  & 4, 5  \\  
PSS1339+5154  & 4.080 & 2.783 & 4.029  & 5  \\  
PSS1401+4111  & 4.026 & 2.866 & 3.976  & 4, 5	\\  
PSS1403+4126  & 3.862 & 2.866 & 3.813  & 4, 5	\\  
PSS1418+4449  & 4.323 & 2.974 & 4.270  & 4, 5  \\  
PSS1430+2828  & 4.306 & 2.811 & 4.253  & 2  \\  
PSS1458+6813  & 4.291 & 2.990 & 4.238  &    \\ 
GB1508+5714   & 4.304 & 3.217 & 4.251  & 2	\\  
PSS1543+3417  & 4.407 & 3.071 & 4.353  & 5  \\  
PSS1615+1803  & 4.010 & 2.783 & 3.960  &   	\\  
PSS1721+3256  & 4.040 & 2.802 & 3.990  & 1, 2  \\   
PSS2154+0335  & 4.359 & 3.400 & 4.305  & 1, 2, 3  \\  
PSS2244+1005  & 4.040 & 2.810 & 3.990  &	\\  
\hline
\multicolumn{5}{l}{$^*$ Quasars from this survey were previously used for $\Omega_{\rm DLA, H_I}$}\\
\multicolumn{5}{l}{estimations by: (1) P\'eroux et al. (2001); (2) P\'eroux et al. (2003) }\\
\multicolumn{5}{l}{(3)P\'eroux et al. (2005); (4) Prochaska et al. (2005); Prochaska \& Wolfe (2009)}\\
\end{tabular}
\end{center}
\end{table}

\begin{table}[h!]
\begin{center}
\caption{Principal absorption metal lines most frequently detected associated to high column density absorption systems.}
\label{princ_lines} 
\vspace{-0.5pc}
\begin{tabular}{l l l l}
\hline
Ion & $\lambda _0$ (\AA)&  \textit{f}~$^a$ & \textit{log$\lambda _0 f  + log[N / N(H)]_{\odot}+12.00$ } \\		 	
\hline
Si~{\sc iii} 	&		1206.500		&	 0.221				&	10.87		\\
H~{\sc i}    	&      	1215,6701		&	 0.4162				&   14.70		\\
N~{\sc v} 		&		1238.821		&	 0.152				&   10.24		\\
N~{\sc v} 		&		1242.804		&	 0.0757				&   9.93		\\
Si~{\sc ii} 	&		1260.4223		&	 0.959				&	10.65		\\
O~{\sc i} 	 	&  		1302,1685     	&    0.0486		      	&   10.67  		\\
Si~{\sc ii} 	&		1304,3702		&	 0.147				&	9.85		\\
C~{\sc ii}  	&  		1334,5323    	&    0.118		       	&   10.85   	\\
Si~{\sc iv} 	&     	1393.76018   	&    0.528	        	&   10.44		\\
Si~{\sc iv}	 	&     	1402.770   		&    0.262	        	&   10.13		\\
Si~{\sc ii} 	&     	1526,70698  	&    0.23			    &   10.11   	\\
C~{\sc iv} 		&     	1548.2041  		&    0.194      		&   11.13		\\
C~{\sc iv} 		&     	1550.7812  		&    0.097      		&   10.83		\\
Fe~{\sc ii} 	&		1608.45085		&	 0.062				&	9.52		\\
Al~{\sc ii} 	&     	1670,7886  		&    1.88			    &   9.99		\\
Al~{\sc iii} 	&     	1854.7164  		&    0.539			    &   9.49		\\
Al~{\sc iii} 	&     	1862.7895  		&    0.268			    &   9.19		\\
Fe~{\sc ii} 	& 		2344.2139		&	 0.108				&	9.92		\\
Fe~{\sc ii}		& 		2374.4162		&	 0.0395				&	9.49		\\
Fe~{\sc ii}		& 		2382.7652		&	 0.328				&	10.41		\\	
\hline\\
\multicolumn{4}{l}{$^a$ Oscillator strengths taken from Morton (1991).}\\
\end{tabular}
\end{center}
\end{table}			

\begin{table*}
\caption{Frequency distribution fitting parameters }
\begin{tabular}{llllll}
\hline\hline
Form      	     & $z_{\rm range}$ & log K            &  $\alpha$             & $log N_T$       & $\beta$        \\
\hline
Power-Law 	     & [2.55 - 5.03]   & 6.59$\pm$2.12    &   $-$1.384$\pm$0.105  &  -------        & -------         \\
	             & [2.55 - 3.40]   & 5.55$\pm$2.88    &   $-$1.330$\pm$0.144  &  -------        & -------         \\
      		     & [3.40 - 3.83]   & 4.03$\pm$2.01    &   $-$1.056$\pm$0.199  &  -------        & -------         \\
      		     & [3.83 - 5.03]   & 4.38$\pm$2.39    &   $-$1.279$\pm$0.118  &  -------        & -------         \\
\hline
Double Power-Law & [2.55 - 5.03]   & -21.78$\pm$0.31  &   $-$1.162$\pm$0.118  &  20.60$\pm$0.24 & -2.07$\pm$0.51  \\
	             & [2.55 - 3.40]   & -22.64$\pm$1.25  &   $-$1.327$\pm$0.307 &  21.19$\pm$1.44 & -5.98$\pm$2.07  \\
      		     & [3.40 - 3.83]   & -21.28$\pm$0.19  &   $-$0.793$\pm$0.233  &  20.45$\pm$0.14 & -2.49$\pm$0.74  \\
      		     & [3.83 - 5.03]   & -21.63$\pm$0.25  &   $-$0.920$\pm$0.199  &  20.51$\pm$0.17 & -2.10$\pm$0.44  \\
\hline
Gamma	  	     & [2.55 - 5.03]   & -21.95$\pm$0.36  &   $-$1.010$\pm$0.172  &  20.93$\pm$0.18 & -------         \\
  		         & [2.55 - 3.40]   & -22.53$\pm$1.54  &   $-$1.185$\pm$0.329  &  21.28$\pm$0.89 & -------         \\
      		     & [3.40 - 3.83]   & -20.96$\pm$0.28  &   $-$0.485$\pm$0.324  &  20.45$\pm$0.16 & -------         \\
      		     & [3.83 - 5.03]   & -21.80$\pm$0.35  &   $-$0.853$\pm$0.198  &  20.82$\pm$0.18 & -------         \\
\hline
\label{par_freq}
\end{tabular}
\end{table*}

\begin{table*}
\caption{Absorption Distance Path - Data for Figure \ref{omega_dla}}
\begin{tabular}{llllllll}
\hline\hline
$z_{\rm range}$    &  $<z>$   & $N_{\rm QSO}$ & $N_{\rm DLA}$ & $N_{\rm subDLA}$  & $\Delta X$  & $\Omega_{\rm DLA} (10^3)$  & $\Omega_{\rm DLA+subDLA} (10^3)$\\
\hline
2.55 - 3.40    &  3.168   &   77      &   12      &     23         &  125.922   &  $1.43 \pm 0.33$        &  $1.71 \pm 0.33$  \\

3.40  - 3.83   &  3.618   &   78      &   17      &     19         &  125.922   &  $1.41 \pm 0.26$        &  $1.65 \pm 0.26$  \\

3.83 - 5.03    &  4.048   &   77      &   11      &     18         &  125.922   &  $0.97 \pm 0.22$        &  $1.21 \pm 0.22$   \\
\hline
\label{omegaHI}
\end{tabular}
\end{table*}

\newpage

\onecolumn
\begin{appendix}

\begin{center}
\bf{Appendix : Notes on individual systems}
\end{center}

In this Appendix we discuss systems where we note differences between our measurements
and measurements by others.


\begin{enumerate}

\item PSS~0131+0633 ($z_{\rm em}$= 4.430). P\'eroux et al. (2001) report two sub-DLA candidates 
at $z_{\rm abs}$= 3.17 and 3.61 with log~$N$(H~{\sc i}) = 19.9 and 19.8 respectively.
For the second system however, P\'eroux et al. (2001) give in Table~5 a redshift of  $z_{\rm abs}$~=~3.61 for the 
H~{\sc i} line and $z_{\rm abs}$~=~3.609 for the metal lines but a redshift of $z_{\rm abs}$~=~3.69 and
log~$N$(H~{\sc i}) = 19.5 in their Section 8 namely, "Notes on Individual Objects". 
We also detect the first sub-DLA at $z_{\rm abs}$~=~3.173 with log$N$(H~{\sc i}) = 19.95 $\pm$ 0.10. 
Metal lines at that redshift are detected in the red part of the spectrum. 
For the second DLA candidate, we confirm the presence of the $z_{\rm abs}$~=~3.689 absorption with 
log~$N$(H~{\sc i}) =  19.50 $\pm$ 0.10 in agreement with the sub-DLA 
candidate reported in Section 8 of Peroux et al. (2001). Metal lines at that redshift are also 
observed in the red part of the spectrum.

\item PSS0747+4434 ($z_{\rm em}$~=~4.435). P\'eroux et al. (2001) report two DLA candidates at 
$z_{_rm abs}$~=~3.76 and 4.02 with log~$N$(H~{\sc i}) = 20.3 and 20.6 respectively, which are confirmed 
by our observations. We measure log~$N$(H~{\sc i}) = 20.00~$\pm$~0.20 at $z_{\rm abs}$~=~3.762 and 
log~$N$(H~{\sc i}) = 21.10~$\pm$~0.15 at $z_{\rm abs}$~=~4.019. No metal lines at $z_{\rm abs}$~=~3.76 
are observed in both surveys. We remind the reader that our data are of higher spectral resolution.

\item SDSS~0756+4104 ($z_{\rm em}$~=~5.09). We detect a sub-DLA candidate at $z_{\rm abs}$~=~4.360 with 
log~$N$(H~{\sc i}) = 20.15~$\pm$~0.10. Although no Lyman series and metals are detected for this system, the 
spectrum has a high enough SNR to fit the H~{\sc i} line well and to ascertain that this aborber is highly likely to 
be Damped. 

\item BR~0951$-$450 ($z_{\rm em}$~=~4.35). PMSI03 report the detection of two DLA candidates at $z_{\rm abs}$~=~3.8580 
and 4.2028 with log~$N$(H~{\sc i}) = 20.6 and 20.4 respectively, which are confirmed by our observations. 
We measure log~$N$(H~{\sc i}) = 20.70~$\pm$~0.20 at $z_{\rm abs}$~=~3.856 and log~$N$(H~{\sc i}) = 20.35~$\pm$~0.15 
at $z_{\rm abs}$~=~4.202. We discover an additional Damped candidate at $z_{\rm abs}$~=~3.235 with 
log~$N$(H~{\sc i}) = 20.25~$\pm$~0.10. We believe that this system is not included in the statistical sample of PMSI03, 
because it falls below the threshold of log~$N$(H~{\sc i}) = 20.30. Metal lines at that redshift are observed in the red 
part of the spectrum.

\item BRI~1013+0035 ($z_{\rm em}$~=~4.38). PMSI03 report one DLA candidate at $z_{\rm abs}$~=~3.103 with 
log~$N$(H~{\sc i}) = 21.1. This system is not included in our sample since its redshift is smaller than the 
minimum redshift beyond which a DLA/sub-DLA could be detected in our spectrum. One Damped candidate was 
discovered at $z_{\rm abs}$~=~3.738 with log~$N$(H~{\sc i}) = 20.15~$\pm$~0.20. We believe that this system is 
not included in the statistical sample of PMSI03 because it falls below the threshold of log~$N$(H~{\sc i}) = 20.30. 
Although no lyman series and metals are detected for this system, the SNR in the spectrum is high enough to 
ascertain that this absorber is highly likely to be Damped. 

\item PSS~1057+4555 ($z_{\rm em}$~=~4.126). P\'eroux et al. (2001) report three DLA candidates at $z_{\rm abs}$~=~2.90, 
3.05 and 3.32 with log~$N$(H~{\sc i}) = 20.1, 20.3 and 20.2 respectively, which are confirmed by our observations. 
We measure log~$N$(H~{\sc i}) = 20.05~$\pm$~0.10 at $z_{\rm abs}$~=~ 2.909 , log~$N$(H~{\sc i}) = 19.80~$\pm$~0.15 
at $z_{\rm abs}$~=~ 3.058 and log~$N$(H~{\sc i}) = 20.15~$\pm$~0.10 at $z_{\rm abs}$~=~ 3.317. 
Metal lines at all three redshifts are observed in the red part of the spectrum. We report a new 
sub-DLA at $z_{\rm abs}$~=~3.164 with log~$N$(H~{\sc i}) = 19.50~$\pm$~0.20. 
Several metal lines are detected for this sytem.

\item PSS~1253$-$0228 ($z_{\rm em}$~=~4.007). P\'eroux et al. (2001) report two DLA candidates at 
$z_{\rm abs}$~=~2.78 and 3.60 with log~$N$(H~{\sc i}) = 21.4 and 19.7, respectively. The system at 
$z_{\rm abs}$~=~ 2.78 is not included in our sample because its redshift falls below the minimum redshift  
beyond which DLA/sub-DLA could be detected in our spectrum. We measure log~$N$(H~{\sc i}) = 19.95~$\pm$~0.10 at 
$z_{\rm abs}$~=~ 3.608. Several metal lines are detected for this sytem.

\item J~1325+1123 ($z_{\rm em}$~=~4.400). We have discovered two sub-DLA candidates at $z_{\rm abs}$~=~3.723 
and 4.133 with log~$N$(H~{\sc i}) = 19.50~$\pm$~0.20 and 19.50~$\pm$~0.20, respectively. C~{\sc iv} metal lines 
are detected at $z_{\rm abs}$~=~3.723. This two sub-DLAs can be found in the list of SDSS DR5 systems available 
on Jason Prochaska's website.  

\item PSS~1633+1411 ($z_{_rm em}$~=~4.349). P\'eroux et al. (2001) report one sub-DLA candidate at 
$z_{\rm abs}$~=~3.90 with log~$N$(H~{\sc i}) = 19.8. Here, we measure log~$N$(H~{\sc i}) = 19.0~$\pm$~0.20 at 
$z_{\rm abs}$~=~ 3.909. We report a new DLA at $z_{\rm abs}$~=~2.880 with log~$N$(H~{\sc i}) = 20.30~$\pm$~0.15. 
Several metal lines are detected for this sytem.

\item PSS~1646+5514 ($z_{\rm em}$~=~4.084). No DLA candidate is detected by P\'eroux et al. (2001) along this line
of sight. We report here two sub-DLA candidates at $z_{\rm abs}$~=~ 2.932 and 4.029 with 
log~$N$(H~{\sc i}) = 19.50~$\pm$~0.10 and 19.80~$\pm$~0.15, respectively. No metals are seen
over the available wavelength coverage. 

\item PSS~2122$-$0014 ($z_{\rm em}$~=~4.084). P\'eroux et al. (2001) report two candidates at 
$z_{\rm abs}$~=~3.20 and 4.00 with log~$N$(H~{\sc i}) = 20.3 and 20.1, respectively. Here, we 
measure log~$N$(H~{\sc i}) = 20.20~$\pm$~0.10 and 20.15~$\pm$~0.15 at $z_{\rm abs}$~=~ 3.207 and 4.001. 
Several metal lines are detected for both systems. We report a new sub-DLA candidate at $z_{\rm abs}$~=~3.264 
with log~$N$(H~{\sc i}) = 19.90~$\pm$~0.15. Several metal lines are detected also for this system. 

\item PSS~2154+0335 ($z_{\rm em}$~=~4.359). P\'eroux et al. (2001) report two candidates at $z_{\rm abs}$~=~3.61 
and 3.79 with log~$N$(H~{\sc i}) = 20.4 and 19.70, respectively. No candidate with log~$N$(H~{\sc i})~$>$~19.50 
is detected in our spectrum.

\item PSS~2155+1358 ($z_{\rm em}$= 4.256). P\'eroux et al. (2001) report one DLA at $z_{\rm abs}$~=~3.32 
with log~$N$(H~{\sc i}) = 21.10. Dessauges et al. (2003) report three sub-DLAs at $z_{\rm abs}$~=~3.142, 
3.565 and 4.212 with log~$N$(H~{\sc i}) = 19.94, 19.37 and 19.61. Here, we measure 
log~$N$(H~{\sc i}) = 19.90~$\pm$~0.20, 20.75~$\pm$~0.20 and 20.00~$\pm$~0.10 at $z_{rm abs}$~=~3.143, 3.318 and 
4.211, respectively. Several metal lines are detected for these systems.

\item PSS2344+0342 ($z_{\rm em}$~=~4.340). P\'eroux et al. (2001) report two DLAs at $z_{\rm abs}$~=~2.68 
and 3.21 with log~$N$(H~{\sc i}) = 21.10 and 20.9. The first system is not included in our sample because
its redshift falls below the minimum redshift beyond which a DLA/sub-DLA could be detected in our spectrum. 
We measure log~$N$(H~{\sc i}) = 21.20~$\pm$~0.10 at $z_{\rm abs}$~=~ 3.220. 
Dessauges et al. (2003) report one sub-DLA at $z_{\rm abs}$~=~ 3.882 with log~$N$(H~{\sc i}) = 19.50. 
We measure log~$N$(H~{\sc i}) = 19.80~$\pm$~0.10 at $z_{\rm abs}$~=~3.884.

\end{enumerate}

\newpage

\begin{center}
\bf{Appendix : List of Figures}
\end{center}
\begin{figure*}[b!]
  \includegraphics[width=18cm,height=21.5cm]{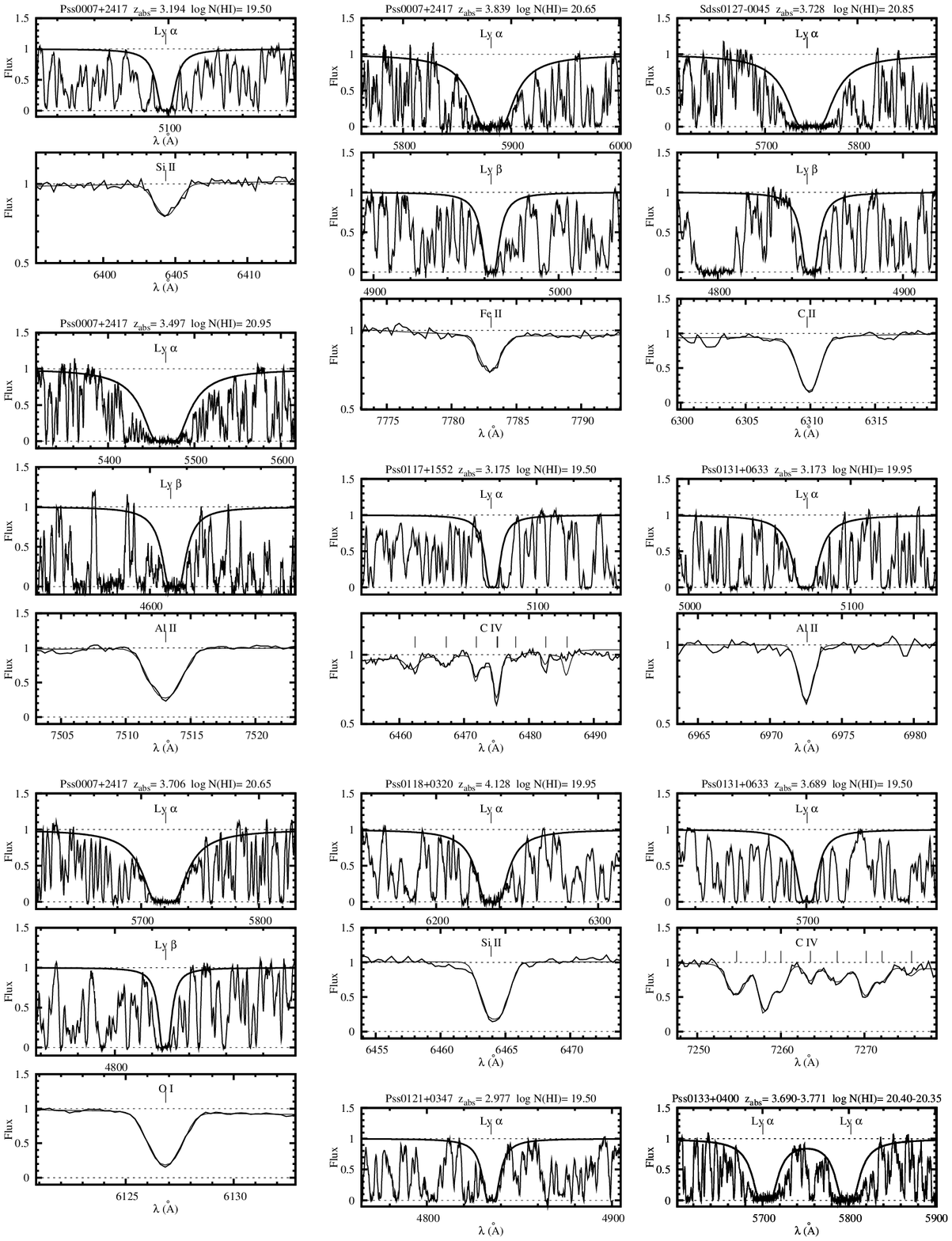}
  \caption{Spectra of the 100 confirmed Lyman-$\alpha$ absorption with log $N$(H~{\sc i}) $\geq$ 19.5.
   In each panel the Voigt profile fit corresponding to the best $N$(H~{\sc i}) value
   is overplotted. The Lyman series absorption lines and one characteristic associated metal absorption 
feature (when metals are present over the available wavelength coverage and not blended with another lines) 
are presented in additional sub-panels.
}
\end{figure*}

\begin{figure*}
  \addtocounter{figure}{-1}
  \includegraphics[width=\textwidth]{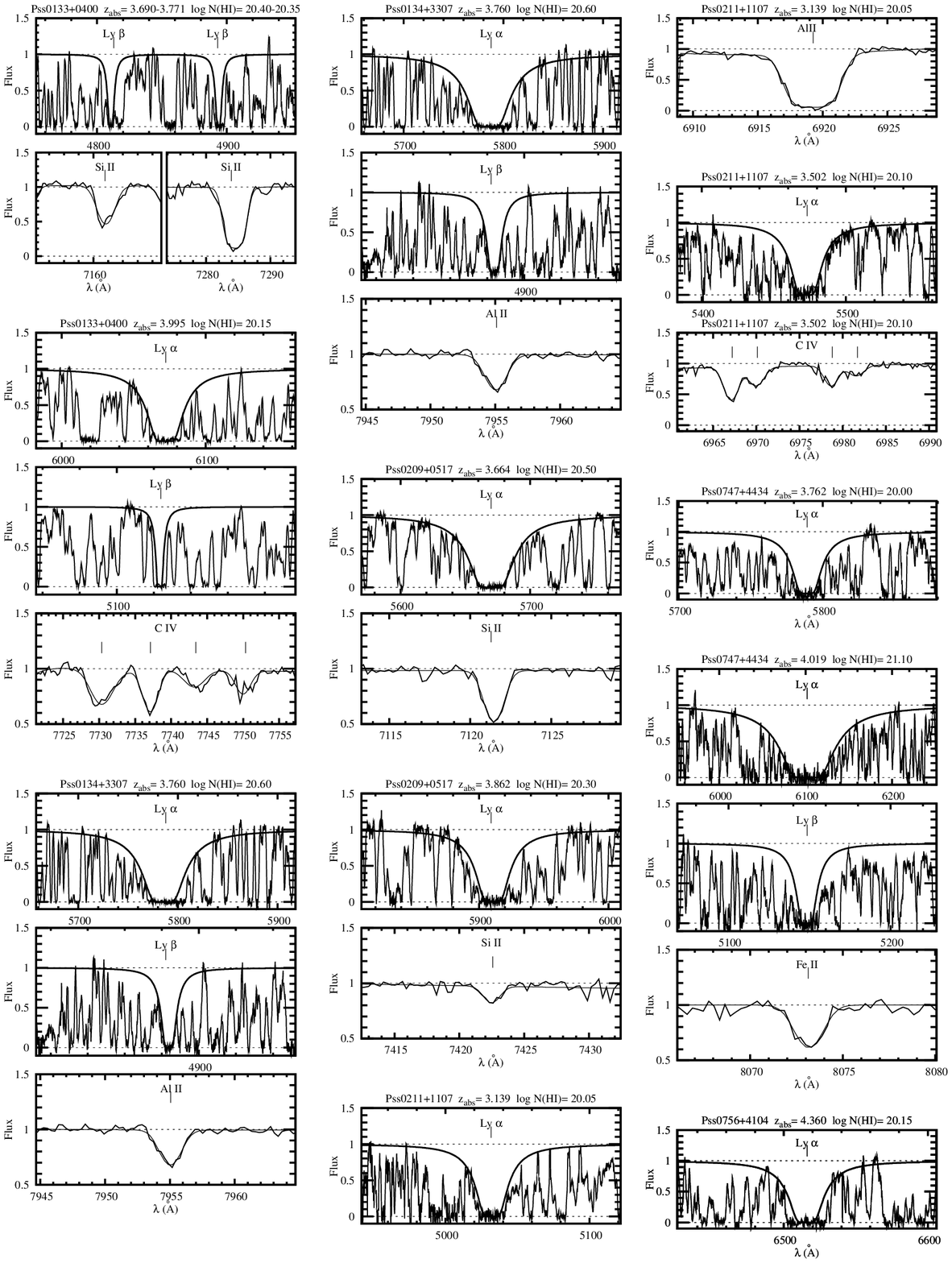}
  \caption{Continued.}
\end{figure*}

\begin{figure*}
  \addtocounter{figure}{-1}
  \includegraphics[width=\textwidth]{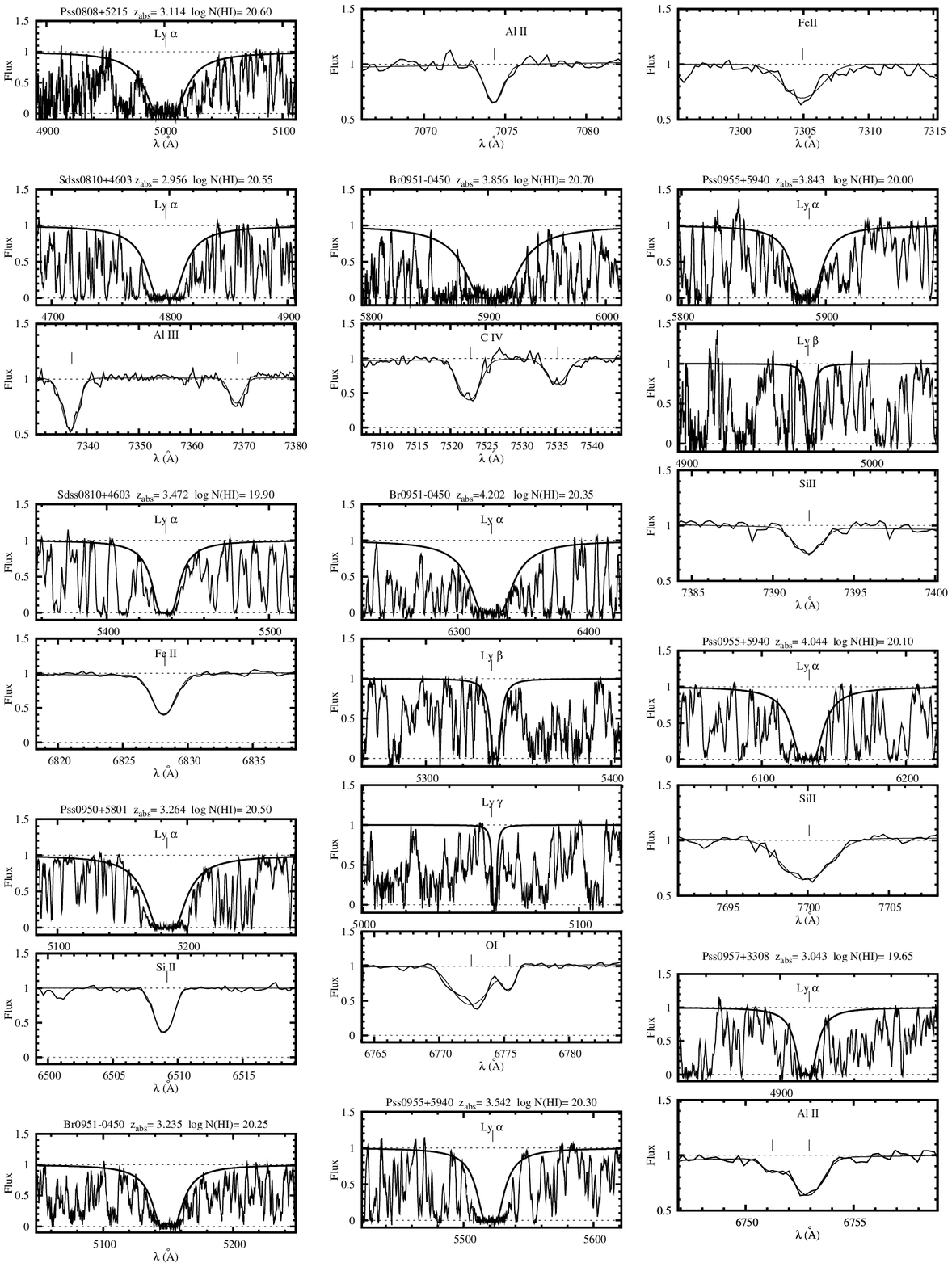}
  \caption{Continued.}
\end{figure*}

\begin{figure*}
  \addtocounter{figure}{-1}
  \includegraphics[width=\textwidth]{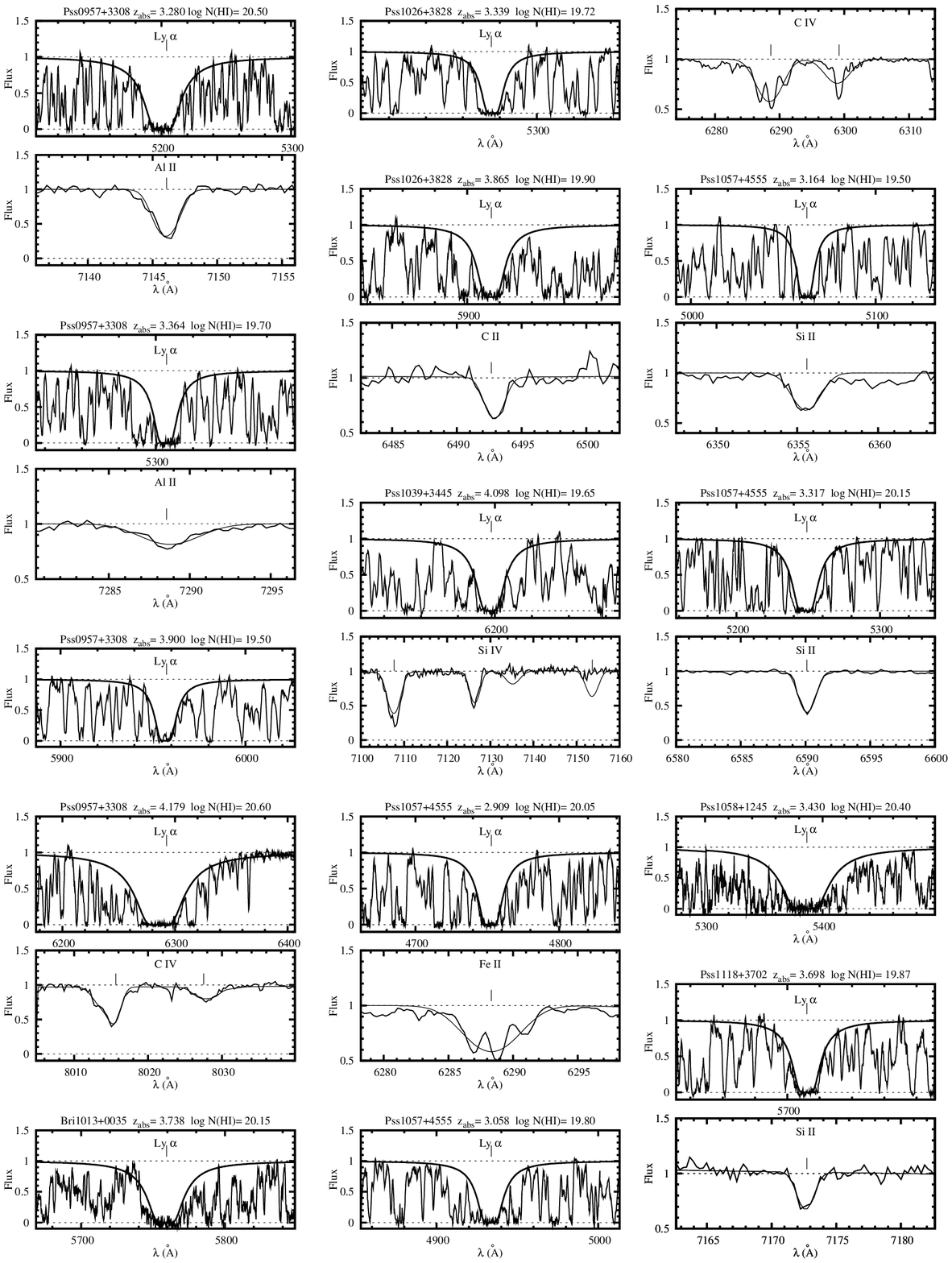}
  \caption{Continued}
\end{figure*}

\begin{figure*}
  \addtocounter{figure}{-1}
  \includegraphics[width=\textwidth]{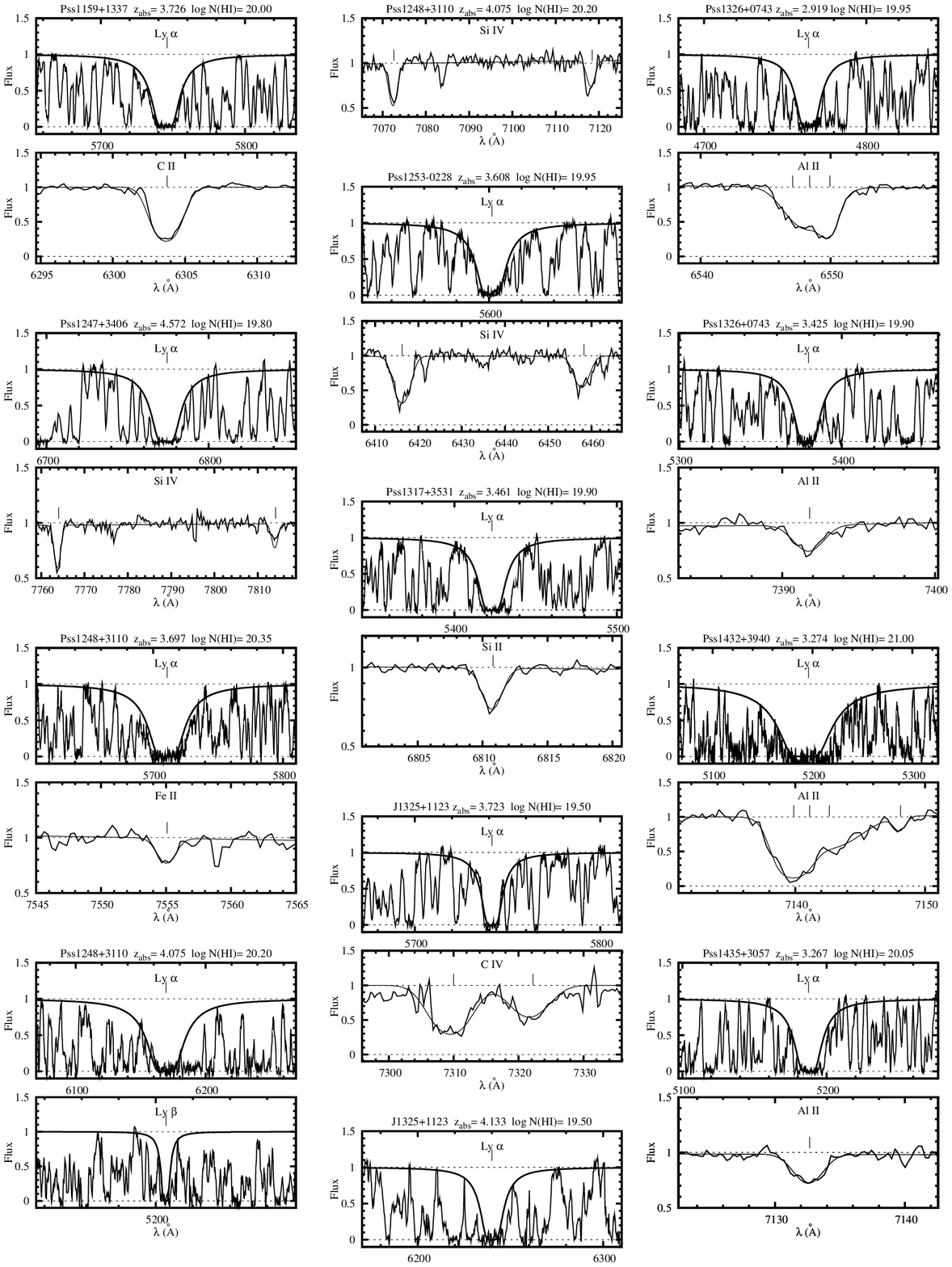}
  \caption{Continued}
\end{figure*}

\begin{figure*}
  \addtocounter{figure}{-1}
  \includegraphics[width=\textwidth]{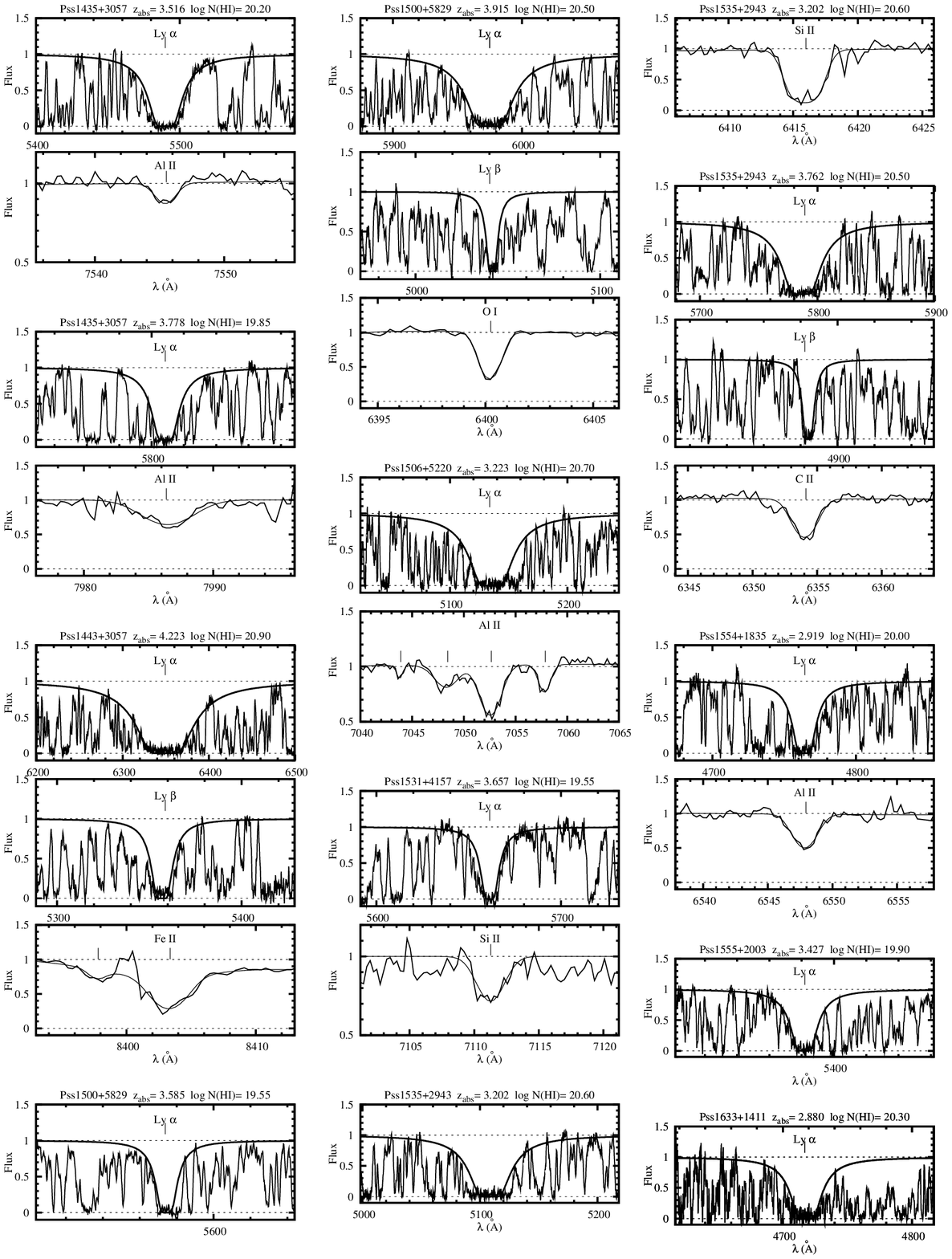}
  \caption{Continued}
\end{figure*}

\begin{figure*}
  \addtocounter{figure}{-1}
  \includegraphics[width=\textwidth]{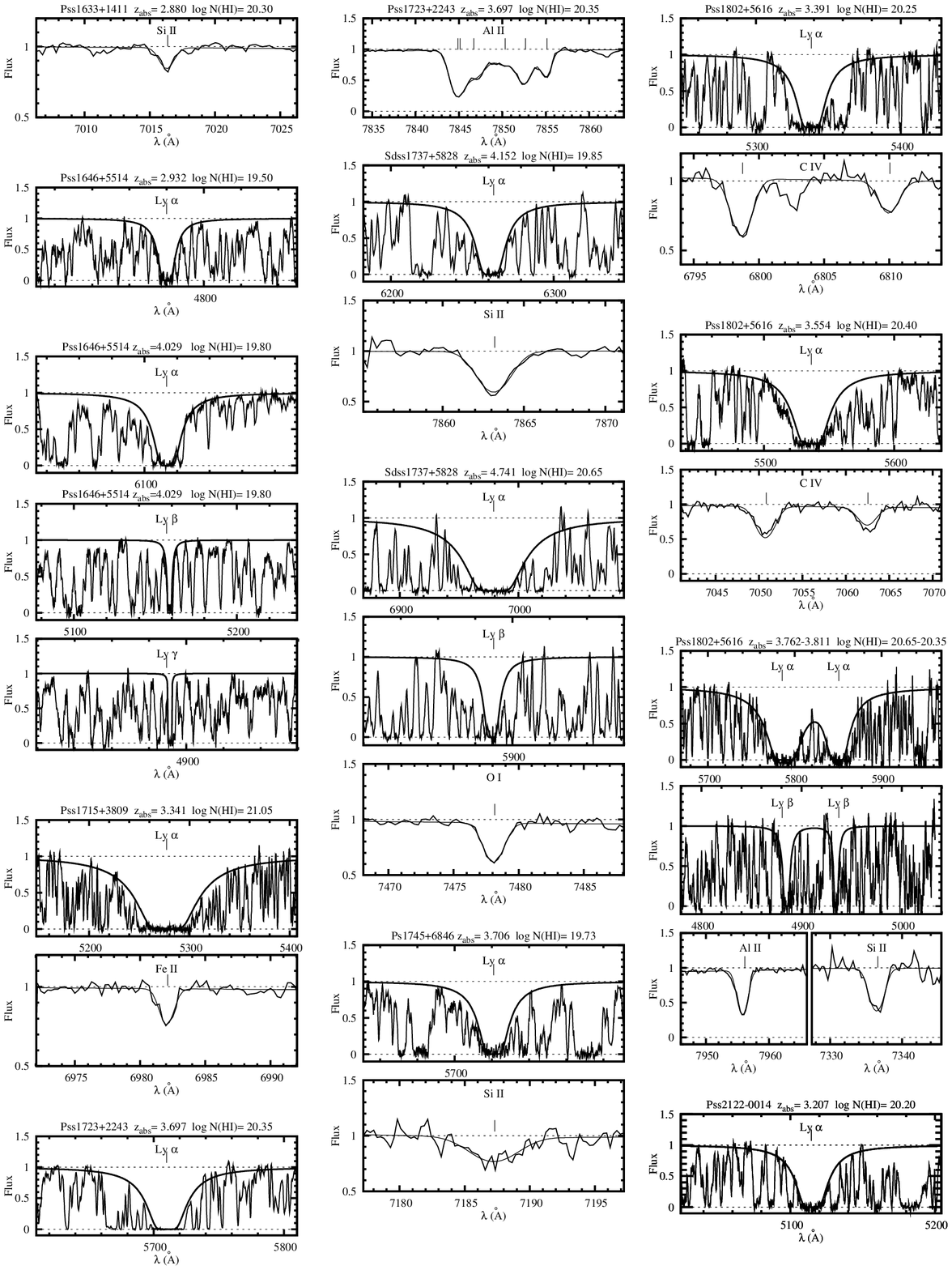}
  \caption{Continued}
\end{figure*}

\begin{figure*}
  \addtocounter{figure}{-1}
  \includegraphics[width=\textwidth]{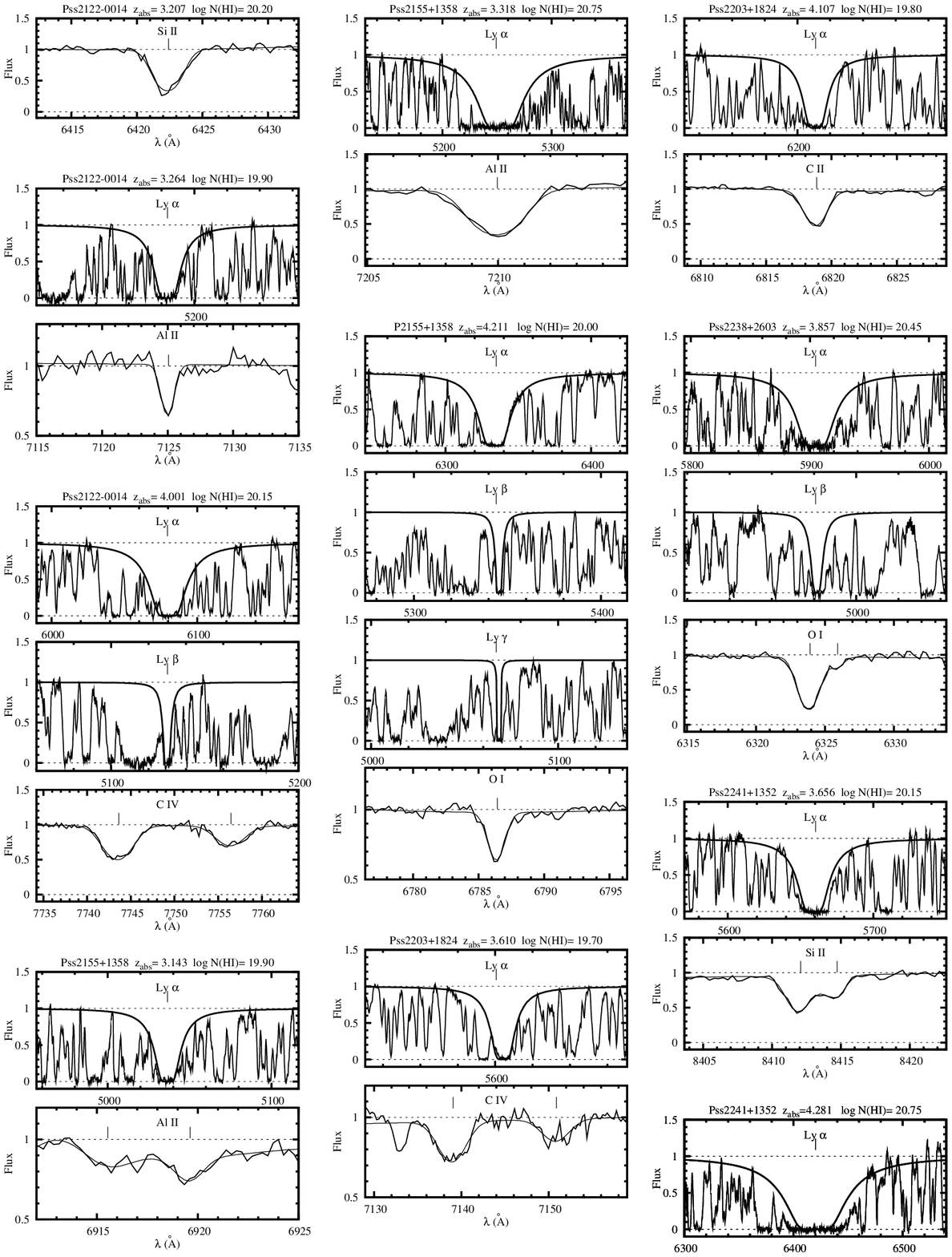}
  \caption{Continued}
\end{figure*}

\begin{figure*}
  \addtocounter{figure}{-1}
  \includegraphics[width=\textwidth]{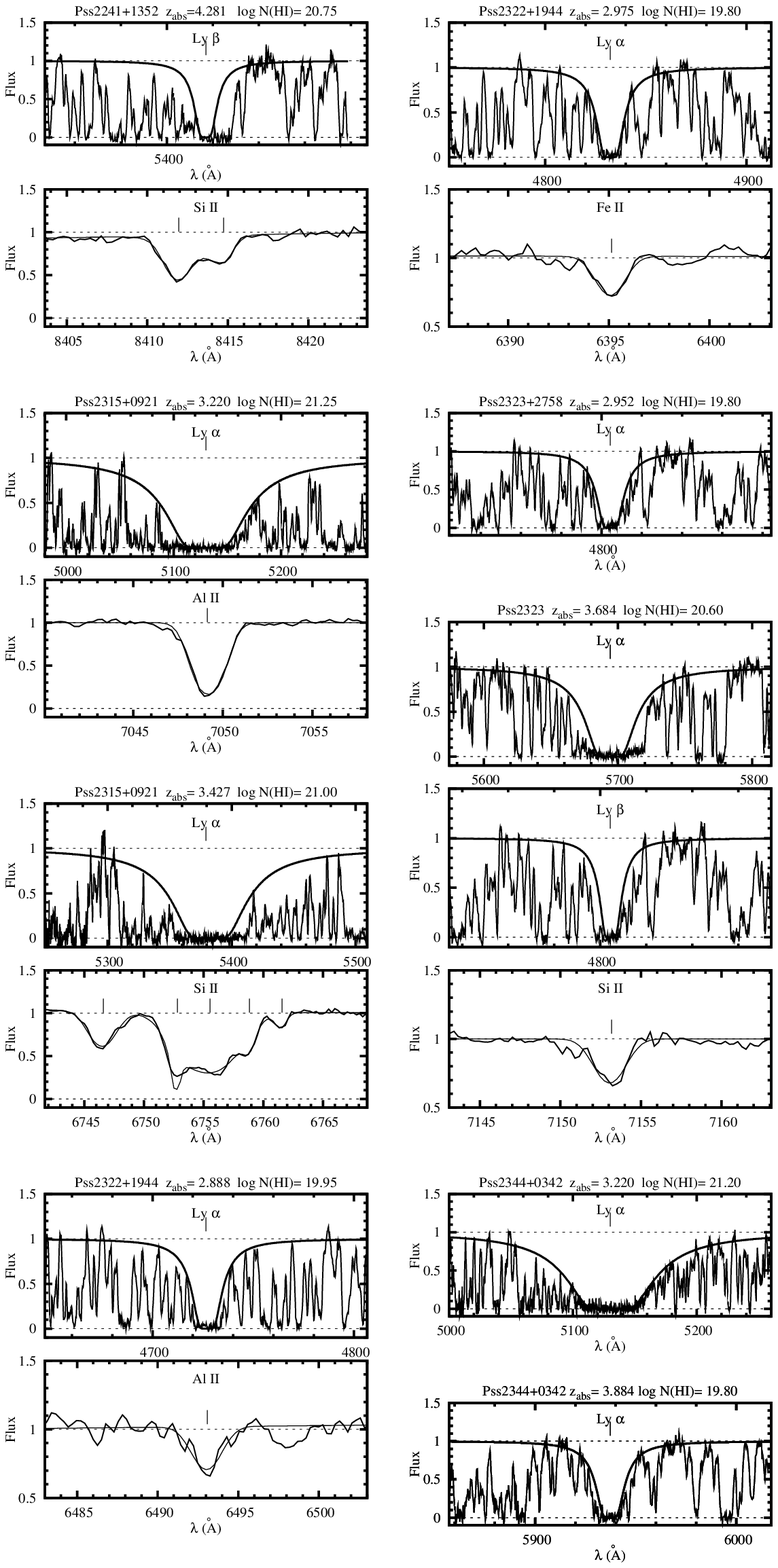}
  \caption{Continued}
\end{figure*}

\end{appendix}

\end{document}